\documentclass[sigconf]{acmart} 

\usepackage{tikz}
\usepackage{bm}
\tikzset{global scale/.style={
		scale=#1,
		every node/.append style={scale=#1}
	}
}
\usepackage{multirow}
\usepackage{mathrsfs}
\usepackage{threeparttable}

\newtheorem{definition}{Definition}

\AtBeginDocument{%
  \providecommand\BibTeX{{%
    \normalfont B\kern-0.5em{\scshape i\kern-0.25em b}\kern-0.8em\TeX}}}

\setcopyright{acmcopyright}
\copyrightyear{2020}
\acmYear{2020}
\setcopyright{acmcopyright}\acmConference[JCDL '20]{Proceedings of the ACM/IEEE
	Joint Conference on Digital Libraries in 2020}{August 1--5, 2020}{Virtual Event, China}
\acmBooktitle{Proceedings of the ACM/IEEE Joint Conference on Digital Libraries in
	2020 (JCDL '20), August 1--5, 2020, Virtual Event, China}
\acmPrice{15.00}
\acmDOI{10.1145/3383583.3398518}
\acmISBN{978-1-4503-7585-6/20/06}

\begin{document}
\fancyhead{}
\title{Multivariate Relations Aggregation Learning in Social Networks}

\author{Jin Xu$^{1}$, Shuo Yu$^{1}$, Ke Sun$^{1}$, Jing Ren$^{1}$, Ivan Lee$^{2}$, Shirui Pan$^{3}$, Feng Xia$^{4}$}
\affiliation{%
	\department{$^{1}$ School of Software, Dalian University of Technology, Dalian, China \\ $^{2}$ School of IT and Mathematical Sciences, University of South Australia, Adelaide, Australia \\ $^{3}$ Faculty of Information Technology, Monash University, Melbourne, Australia \\ $^{4}$ School of Science, Engineering and Information Technology, Federation University Australia, Ballarat, Australia}}
\email{xujin0909@hotmail.com, y\_shuo@outlook.com, kern.sun@outlook.com, ch.yum@outlook.com}
\email{ivan.lee@unisa.edu.au, shirui.pan@monash.edu, f.xia@ieee.org}

\renewcommand{\shortauthors}{J. Xu et al.}

\begin{abstract}
Multivariate relations are general in various types of networks, such as biological networks, social networks, transportation networks, and academic networks. Due to the principle of ternary closures and the trend of group formation, the multivariate relationships in social networks are complex and rich. Therefore, in graph learning tasks of social networks, the identification and utilization of multivariate relationship information are more important. Existing graph learning methods are based on the neighborhood information diffusion mechanism, which often leads to partial omission or even lack of multivariate relationship information, and ultimately affects the accuracy and execution efficiency of the task. To address these challenges, this paper proposes the multivariate relationship aggregation learning (MORE) method, which can effectively capture the multivariate relationship information in the network environment. By aggregating node attribute features and structural features, MORE achieves higher accuracy and faster convergence speed. We conducted experiments on one citation network and five social networks. The experimental results show that the MORE model has higher accuracy than the GCN (Graph Convolutional Network) model in node classification tasks, and can significantly reduce time cost.
\end{abstract}



\maketitle

\section{Introduction}
\label{Section:1}

We are living in a world full of relations. It is of great significance to explore existing social relationships as well as predict potential relationships. These abundant relations generally exist among multiple entities; thus, these relations are also called multivariate relations. Multivariate relations are the most fundamental relations, containing interpersonal relations, public relations, logical relations, social relations, etc~\cite{white2016pluralized, chang2019Multivariat}. Multivariate relations are of more complicated structures comparing with binary relations. Such higher-order structures contain more information to express inner relations among multiple entities. Since multivariate relations can maintain a larger volume of knowledge, multivariate relations are widely applied in a variety of application scenarios, such as anomaly detection, fraud detection, social and academic network analysis, interaction extraction, digital library system, etc~\cite{xu2018research, nguyen2019fake, Yu2019Academic}. Indeed, in social networks, multivariate relations are of more significance. Scholars have gradually realized the importance of multivariate relations in social networks. There are some previous works focus on closures in social networks and some studies focus on teams, groups, communities, etc~\cite{Yu2017Team, zhang2018Multivariate, Wang2019Sustainable}. All these related studies generally focus on mining some certain patterns, formulating corresponding models, or building some theoretical conclusions, etc.

\begin{figure}[!b]
	\centerline{\includegraphics[scale=0.21]{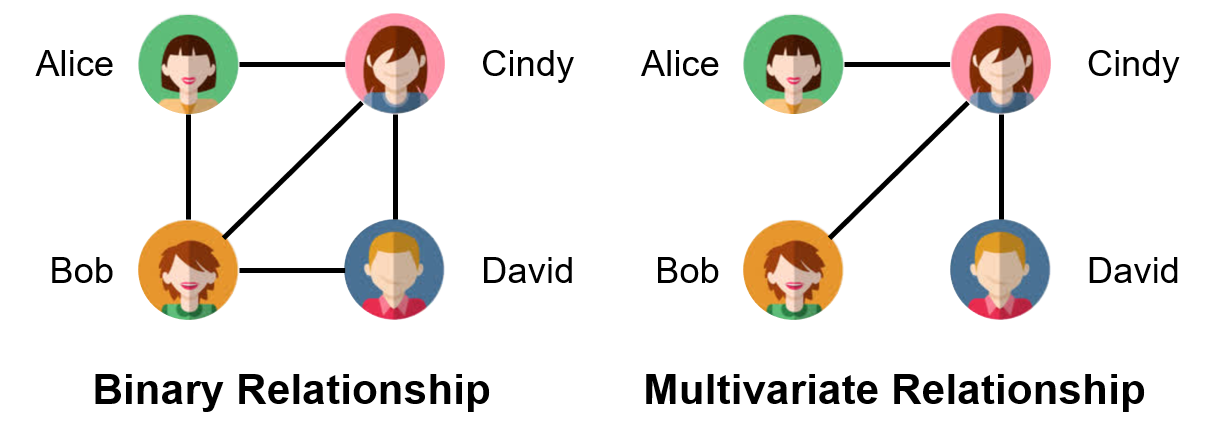}}
	\caption{The Binary and Multivariate Relationship in the Collaboration Network.}
	\label{Fig:showM}
\end{figure}

In some specific scenarios, it seems that multivariate relations can be precisely divided into multiple binary relations. For example, Alice, Bob, Cindy, and David are in one same collaboration team (Figure~\ref{Fig:showM}). It can be depicted by a binary relation with two collaborated members. However, binary relations can only represent simple pairwise relationships, while in most cases, multivariate relations cannot be divided in this way. If Cindy is the team leader, Alice, Bob, as well as David never collaborate, then using binary relations will leading to over-expression or wrong expression of relations. Especially in social networks, multivariate relationships are more common such as friendships, kinships, peer relationships, etc.~\cite{Litwin2016Social}
Such complicated relationships cannot only be represented with binary relationships.

Graph neural network (GNN) has resoundingly taken neighborhood information of nodes into consideration, which is also proved to be more precise and effective~\cite{QIAO2018Data, GOYAL2018Graph}. A certain node may have multiple neighbors, in which some of these neighbors are in closer relations because they are in multivariate relations. However, current GNN methods ignore the relations among neighbors. Based on the principle of ternary closure, there will be a large number of low-order fully-connected structures in interpersonal social networks. This causes complex relationships to appear frequently in social networks, such as network motifs~\cite{Yu2019Motifs}.

In this work, we propose a \textbf{M}ultivariate relati\textbf{O}nship agg\textbf{R}egation l\textbf{E}arning method called \textbf{MORE}. In our proposed method, we employ network motifs to model multivariate relationships in networks, because such the network structure has been proved to be effective in network embedding methods~\cite{rossi2018higher, Rossi2018Deep}.

MORE considers both node attribute features and structural features to enhance the ability of multivariate relation aggregation. For both features, MORE first respectively generates attribute feature tensor and structural feature tensor. Then, by network representation learning, MORE can achieve two embedding tensors (i.e., attribute feature embedding tensor and structural feature tensor) with the same dimension. After then, MORE aggregates two embedding tensors into one aggregated embedding tensor with three different aggregators, including Hadamard aggregator, Summation aggregator, and connection aggregator. Finally, MORE employs softmax to achieve the target vector. Comparing with the baseline method, MORE outperforms with higher accuracy and efficiency in 6 different networks including Cora, Email-Eucore, Facebook, Ploblogs, Football, and TerrorAttack. We specifically analyze that in social networks, our proposed MORE achieves higher accuracy and consumes much less training time. Meanwhile, it is also found that MORE also outperforms the baseline method in the Cora dataset. Generally, we summarize our contributions as follows.

\begin{itemize}
	 \item MORE aggregates both nodes' attribute features and structural features, which is proved to have the ability to better representing multivariate relationships social networks.
	\item MORE consumes much less training time compared with baseline methods. We find that MORE always iterates fewer times than the baseline method and the training time can be shortened to 19.5\% times.
	\item By implementing MORE on 5 different social networks and 1 general dataset, MORE achieves better performance with higher accuracy and efficiency.
\end{itemize}

In this paper, we first introduce the existing graph learning methods and their advantages and disadvantages in Section~\ref{Section:2}. Then, we introduce the design theory and framework of the MORE model in Section~\ref{Section:3}. Next, in Section~\ref{Section:4}, the MORE model will complete the node classification task in six datasets. And MORE will compare with GCN (Graph Convolutional Network) in terms of accuracy and efficiency. Finally, we will discuss the future improvement of the MORE model in Section~\ref{Section:5}. Section~\ref{Section:6} concludes the paper.

\section{Related Work}
\label{Section:2}

\subsection{Digital Library}

The emergence of the Internet and the development of related technologies have not only increased information but also changed the nature of traditional libraries and information services. The Digital Library (DL) has become an important part of the modern digital information system. In a narrow sense, the digital library refers to a library that uses digital technology to process and store various documents; in a broad sense, it is a large-scale, knowledge-free knowledge center that can implement multiple information search and processing functions without time and space restrictions. Academic data sets and online academic literature search platforms, such as IEEE / IEE Electronic Library, Wiley Online Library, Springer LINK, Google Academic Search, etc., can be regarded as representative modern digital libraries.

The knowledge information of digital libraries often need to sort and repair. Due to the large number of documents or the loss of knowledge information, manual repair is often inefficient and inaccurate. To this end, based on existing knowledge information, such as the author of the document, keywords, citation data and other information, efficient and automated realization of knowledge or document classification and missing information prediction has become one of the important topics in the field of digital library research. In the process of solving this research topic, machine learning technology play an important role. Wu et al.~\cite{wu2015citeseerx} have developed CiteSeerX, a digital library search engine, over 5-6 years. They combined traditional machine learning, metadata and other technologies to achieve document classification and deduplication, document and citation clustering, and automatic metadata extraction. Vorgia et al. ~\cite{Vorgia2017Hypatia} used many traditional machine learning methods, including 14 classifiers such as Naive Bayes, SVM, Random Forest, etc., to classify documents based on literature abstracts, and achieved good classification results on their datasets. However, the existing method requires a large number of training samples, to obtain a better model through a long iterative learning process. Meanwhile, because it is based on traditional scatter tables or binary relations, it does not take into account the multivariate relations between knowledge or literature, which limits the further optimization of its task accuracy.

\subsection{Graph Learning \& Classification}

Classification task is one of the most classic application scenarios in the field of machine learning. Many researchers have studied ``how to classify data'' and have created a variety of classification algorithms. Among these algorithms, the logistic regression (LR) algorithm and the support vector machine (SVM) algorithm~\cite{Singh2016review} are the most famous. These algorithms are usually based on mathematical knowledge and optimization theory, and they are simple to implement. By repeatedly performing gradient descent on a unit, the algorithm can obtain very effective classification results~\cite{Kurt2008Comparing, Caigny2018new}.
However, these algorithms have huge flaws: They only consider the attributes of the nodes during the iteration process, and cannot consider the dependencies between the data. In the network data with rich relationships between data nodes, the classification accuracy of such algorithms is not high, and their training efficiency is usually low.

Over time, networks have become the focus of researchers' attention. Machine learning in complex graphs or networks is becoming the focus of research by artificial intelligence scientists. Because traditional machine learning methods cannot effectively combine the related information in the network environment, to improve the efficiency and accuracy of tasks, graph learning algorithms have been proposed to solve graph-related problems. Meanwhile, the great success of the recurrent neural network (RNN)~\cite{chung2014empirical} and the convolutional neural network (CNN)~\cite{Krizhevsky2012ImageNet} in the field of natural language~\cite{Vu2016Bi, kumar2016ask, goldberg2017neural} and computer vision~\cite{Zhang2017Learning, Hershey2017CNN, xiong2019good} have provided new ideas and reference objects for graph learning methods. As a result, graph recurrent neural networks (GRNN) and graph convolutional neural networks (GCNN) have become mainstream methods.

\textbf{Graph Recurrent Neural Network (GRNN).} The graph neural network algorithm suitable for graph structure is first proposed by Gori et al.~\cite{gori2005new}, aiming at improving the traditional neural network (NN) algorithm. Gori et al. use the recursive method to continuously propagate neighbor node information until the node state is stable. This neighborhood information propagation mechanism has become the basis for many subsequent graph learning algorithms. Later, Scarselli et al.~\cite{scarselli2008graph} ameliorated this method. The new algorithm combines the ideas of RNN and applies to many types of graph structures, including directed and undirected graphs, cyclic graphs and acyclic graphs. Li et al.~\cite{li2015gated} combined it with the Gated Recurrent Unit (GRU) theory~\cite{Cho2014GRU}. The resulting Gated Graph Neural Network (GGNN) reduced the number of iterations required, thereby improving the overall efficiency of the learning process. However, on the whole, the GRNN algorithm still costs a lot: From the space perspective, this overhead is reflected in the huge parameter and the intermediate state nodes that need to be stored. From the time perspective, these algorithms require a large number of iterations and long training time.
	
	

\begin{table}[!b]
	\centering
	\caption{The notations and their implications.}
	\label{Table:NT}
	\resizebox{0.45\textwidth}{!}{
		\begin{tabular}{cl}
			\toprule
			\multicolumn{1}{c}{\textbf{Notations}} & \multicolumn{1}{c}{\textbf{Implications}}                       \\
			\midrule
			\multicolumn{2}{l}{\textbf{Basic Operation:}}                                                            \\
			$\bm{|\cdot|}$                         & The number of elements in the set.                              \\
			$\bm{\odot}$                           & Hadamard product operation.                                     \\
			\midrule
			\multicolumn{2}{l}{\textbf{Graph Theory Related:}}                                                               \\
			\textbf{G}                             & A graph.                                                        \\
			\textbf{V}                             & The node set of graph \textbf{G}.                             \\
			\textbf{E}                             & The edge set of graph \textbf{G}.                                        \\
			\textbf{A}                             & The adjacency matrix of graph \textbf{G}.                                \\
			\textbf{D}                             & The degree matrix of graph \textbf{G}.                                   \\
			\midrule
			\multicolumn{2}{l}{\textbf{Graph Learning Method Related:}}                                              \\
			$\bm{I_{a}}$                           & The identity matrix of $\bm{a \times a}$.                              \\
			$\bm{X}$ 								&  The Feature matrix. \\
			$\bm{\bar{A}}$                         & The renormalized adjacency matrix.                   \\
			$\bm{\bar{D}}$                         & The renormalized degree matrix.                      \\
			$\bm{\sigma(\cdot)}$                   & The activation function                                         \\
			$\bm{SM(\cdot)}$                       & The softmax function.                                           \\
			$\bm{W,\Theta}$                        & The weight parameter matrix to be trained.                                \\
			\bottomrule
	\end{tabular}}
\end{table}

\textbf{Graph Convolutional Neural Network (GCNN).} Different methods for defining convolution have created various graph convolutional neural network models. Bruna et al.~\cite{bruna2014spectral} combined spectral graph theory with graph learning methods. They used the graph Fourier transform and Convolution theorem in the field of graph signal processing (GSP)~\cite{Sandryhaila2013Discrete, Perraudin2017Stationary, Ortega2018GSP}, to obtain the graph convolution formula. Defferrard et al.~\cite{Defferrard2016Convolutional} made further improvements on their basis. By cleverly designing convolution kernels and combining Chebyshev polynomials, this algorithm has fewer parameters and can effectively aggregate local features in the graph structure. The GCN model proposed by Kipf et al.~\cite{Kipf2016Semi-Supervised} is the master of this direction. By approximating the Chebyshev polynomial of the first order, their method is more stable while avoiding overfitting. Seo et al.~\cite{Seo2018Structured} combined the long short-term memory (LSTM) mechanism~\cite{sak2014long} in RNN, and proposed the GCRN model that can be applied to dynamic networks. On the other hand, Micheli~\cite{Micheli2009Neural} proposed to directly use the neighborhood information aggregation mechanism to implement the graph convolution operation. Gilmer et al.~\cite{Gilmer2017Neural} integrated this theory and proposed a general framework for this kind of GCNN. Hamilton et al.~\cite{Hamilton2017Inductive} used the sampling method to unify the number of neighbor nodes and established the well-known GraphSage algorithm. Velickovic et al.~\cite{Velickovic2018Graph} combined the attention mechanism~\cite{Chorowski2015Att} with GCNN and proposed the graph attention network (GAT) model. These two ideas have their advantages, but they are still limited to neighborhood information. This will lead to partial omission or even complete loss of the rich multivariate relationship information in the network.

\section{The Design of MORE}
\label{Section:3}

In this section, we introduce the framework of the MORE. First, we illustrate the definition and characteristics of the network motif. Then, the correlation between the multivariate relationship and the network motif will be explained. Finally, we introduce the framework of the MORE model in detail.

\subsection{Network Motif}

The network motif firstly refers to a part that has a specific function or a specific structure in a biological macromolecule. For example, a combination of amino acids that characterize a particular function in a protein macromolecule can be called as a motif. Milo et al.~\cite{Milo2002Motif} extended this concept to the field of network science in 2002. Their research team discovered a frequent subgraph structure while studying gene transcription networks. In subsequent research work, they found some special subgraphs with similar characteristics but different structures in many natural networks. Milo et al. summarized the relevant results and proposed a new network subgraph concept called the network motif.

Unlike network graphlets and communities, the network motif is a special type of subgraph structure in the network, which has a stronger practical scene and significance. By definition, the network motif is a kind of connected induced subgraphs, which have the following three characteristics:

\textbf{(1) Network motifs have stronger practical meaning.} Similar to the concept of motifs in biology, network motifs also have some specific practical significances. This practical meaning is determined by the certain structure of the network motif, the type and the characteristic of the network. For example, a typical three-order triangle motif can represent a three-person friendship in a general social network. Therefore, the number of such network motifs can be effectively used to study the correctness of the ternary closure principle in social networks.
	
\textbf{(2) Network motifs appear more frequently in real world networks.} This phenomenon is caused by the structural characteristics of the network motif and its practical significance. In a real network and a random network with the same number of nodes and edges, the network motif appears much more frequently in the former than the latter. In some real sparse networks, such as transportation networks, some complex network motifs may still occur frequently. But they cannot be found in the corresponding random networks.
	
\textbf{(3) Network motifs are mostly low-order structures.} Network motifs can be regarded as a special kind of low-order subgraph structure. In general, the number of nodes that make up a network motif will not exceed eight. Scholars refer to a network motif consisting of three or four nodes as a low-order network motif, and a network motif composed of five or more nodes is called a high-order network motif. High-order network motifs are numerous, complex, and are often used in the super-large-scale network. Applying it to the analysis of a general network requires a lot of time to perform preprocessing, and it is difficult to obtain better experimental results than using low-order motifs. That is, the input time cost and the output experimental effect often fail to balance. Therefore, in this paper, we mainly use low-order network motifs composed of three or four nodes for network analysis.

\begin{table}[!t]
	\centering
	\caption{Network motifs used in this paper.}
	\label{Table:NM}
	\resizebox{0.39\textwidth}{!}{
		\begin{tabular}{cc|ccccc}
			\toprule
			\textbf{Motif} & \textbf{Code} & $\bm{\bar{d}}$ & $\bm{\max{(d)}}$ & $\bm{D}$ & $\bm{\rho}$ & $\bm{|T|}$ \\
			\midrule
			\begin{tikzpicture}[line width = 1pt, global scale = 0.3,
			A/.style = {circle, draw, fill = black, minimum size = 1mm}]
			\node [A] (A) at (0,0) {};
			\node [A] (B) at (1,0) {};
			\node [A] (C) at (1,1) {};
			\draw (A) -- (B);
			\draw (B) -- (C);
			\draw (A) -- (C);
			\end{tikzpicture}
			& \textbf{M31}  & 2.00 & 2    & 1 & 1.00 & 1                \\
			\begin{tikzpicture}[line width = 1pt, global scale = 0.3,
			A/.style = {circle, draw, fill = black, minimum size = 1mm}]
			\node [A] (A) at (0,0) {};
			\node [A] (B) at (1,0) {};
			\node [A] (C) at (1,1) {};
			\draw (A) -- (B);
			\draw (B) -- (C);
			\end{tikzpicture}
			& \textbf{M32}  & 1.33 & 2    & 2 & 0.67 & 0                \\
			\begin{tikzpicture}[line width = 1pt, global scale = 0.3,
			A/.style = {circle, draw, fill = black, minimum size = 1mm}]
			\node [A] (A) at (0,0) {};
			\node [A] (B) at (1,0) {};
			\node [A] (C) at (0,1) {};
			\node [A] (D) at (1,1) {};
			\draw (A) -- (B);
			\draw (B) -- (C);
			\draw (C) -- (D);
			\draw (D) -- (A);
			\draw (C) -- (A);
			\draw (D) -- (B);
			\end{tikzpicture}
			& \textbf{M41}  & 3.00 & 3    & 1 & 1.00 & 4                \\
			\begin{tikzpicture}[line width = 1pt, global scale = 0.3,
			A/.style = {circle, draw, fill = black, minimum size = 1mm}]
			\node [A] (A) at (0,0) {};
			\node [A] (B) at (1,0) {};
			\node [A] (C) at (0,1) {};
			\node [A] (D) at (1,1) {};
			\draw (C) -- (B);
			\draw (A) -- (C);
			\draw (B) -- (D);
			\draw (C) -- (D);
			\draw (B) -- (A);
			\end{tikzpicture}
			& \textbf{M42}  & 2.50 & 3    & 2 & 0.83 & 2                \\
			\begin{tikzpicture}[line width = 1pt, global scale = 0.3,
			A/.style = {circle, draw, fill = black, minimum size = 1mm}]
			\node [A] (A) at (0,0) {};
			\node [A] (B) at (1,0) {};
			\node [A] (C) at (0,1) {};
			\node [A] (D) at (1,1) {};
			\draw (A) -- (C);
			\draw (B) -- (D);
			\draw (C) -- (D);
			\draw (B) -- (A);
			\end{tikzpicture}
			& \textbf{M43}  & 2.00 & 2    & 2 & 0.67 & 0                \\
			\bottomrule
	\end{tabular}}
\end{table}

At present, the research scope of network motif researchers has expanded from the biochemical network, gene network, and biological neural network to social networks, academic collaboration networks, transportation networks, etc. In these networks, they found a large number of network motifs with different structural characteristics. In this sense, the network motif can reveal the basic structure of most network structures and play an important role in the specific application functions of the network. In Table~\ref{Table:NM}, we list various network motifs used in this paper. Among them, $\bm{\bar{d}}$ denotes the average degree, $\bm{\max{(d)}}$ denotes the maximum degree, $\bm{D}$ denotes the diameter, $\bm{\rho}$ denotes the density, and $\bm{|T|}$ denotes the number of triangles contained in the network motif.

\begin{figure}[!b]
	\centerline{\includegraphics[scale=0.29]{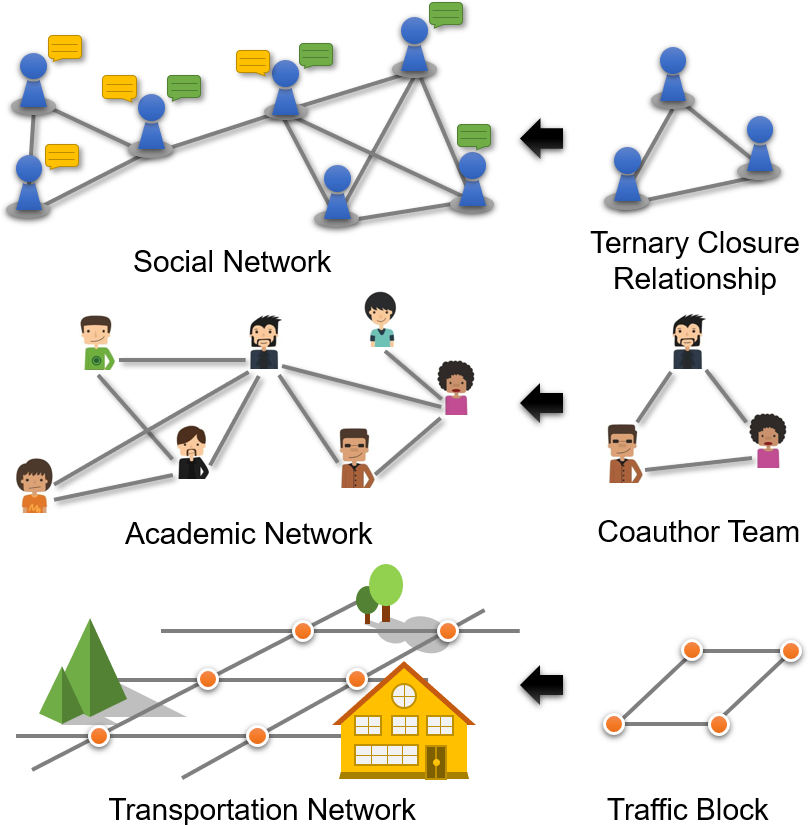}}
	\caption{Schematic diagram of multivariate relationships in a social network, an academic network, and a traffic network.}
	\label{Fig:showMR}
\end{figure}

\subsection{Multivariate Relationship \& Network Motif}

\begin{figure*}[!t]
	\centerline{\includegraphics[scale=0.295]{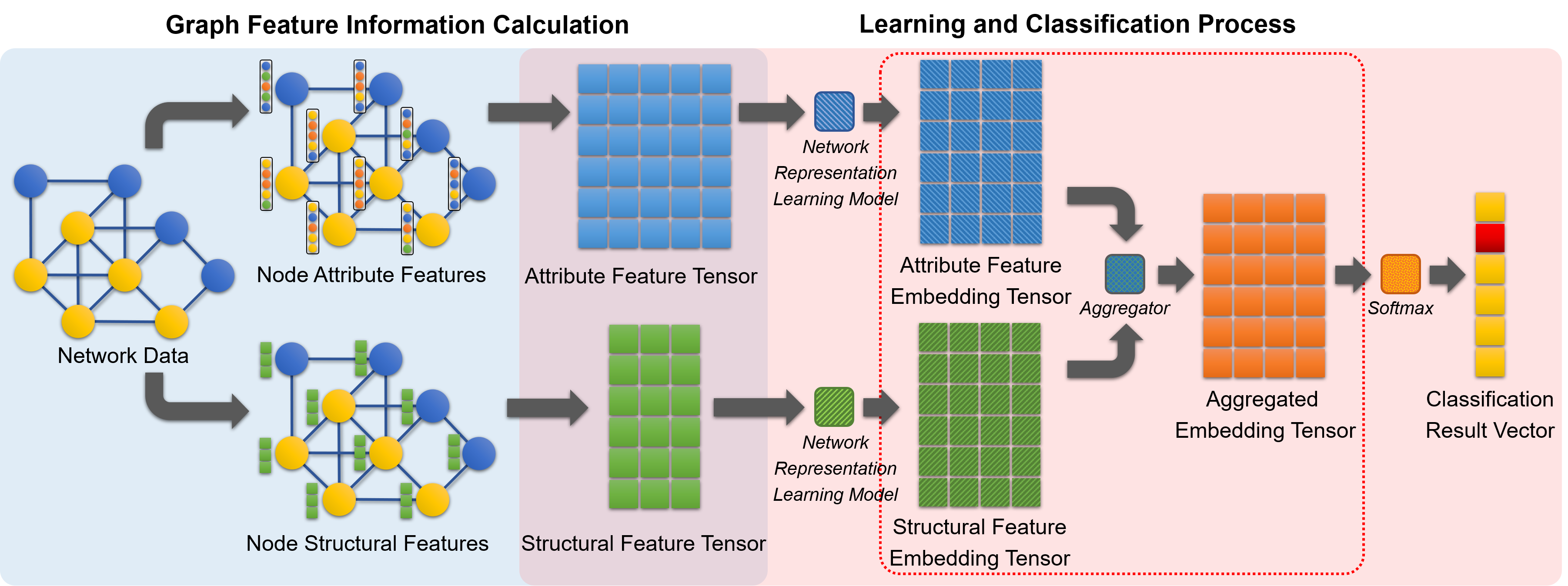}}
	\caption{Schematic diagram of MORE framework in a 2-label network (Represented by blue and yellow nodes in the graph).}
	\label{Fig:Framework}
\end{figure*}

Relationships are the foundation of network building. In the network dataset, the edge between two nodes actually reflects a kind of binary relationship, that is, there is a relationship between the two data entities. This low-order relationship is simple, direct, and easy to characterize. However, because it only considers the association between two data, low-order relationships often lose high-order information within many datasets. To this end, researchers have extended the binary relationship into the multivariate relationship, and analyzed the multivariate relationship to mine patterns, features and implicit associations in the network, to obtain more accurate and valuable conclusions.	

Multivariate relationships are common in actual networks, and the network motif can effectively characterize them because of their definitions and properties. As shown in Figure~\ref{Fig:showMR}, in a social network, besides the direct relationship connect one user to another, the ternary closure relationship between three users is a common phenomenon existing in the social environment. This relationship can be regarded as a fully-connected small group, and fully-connected network motifs, such as \textbf{M31} and \textbf{M41} (in Table~\ref{Table:NM}), can effectively characterize this multivariate relationship in the social network. This kind of network motifs can also be used to characterize the ubiquitous coauthor team consisting of three or four scholars in the academic network. In transportation networks, such as urban road networks, there are a large number of block-like structures in the network, due to the constraints of realistic conditions such as traffic planning and geographical factors. This multivariate relationship between intersections can be effectively characterized by \textbf{M43}.

\begin{figure}[!b]
	\centerline{\includegraphics[scale=0.35]{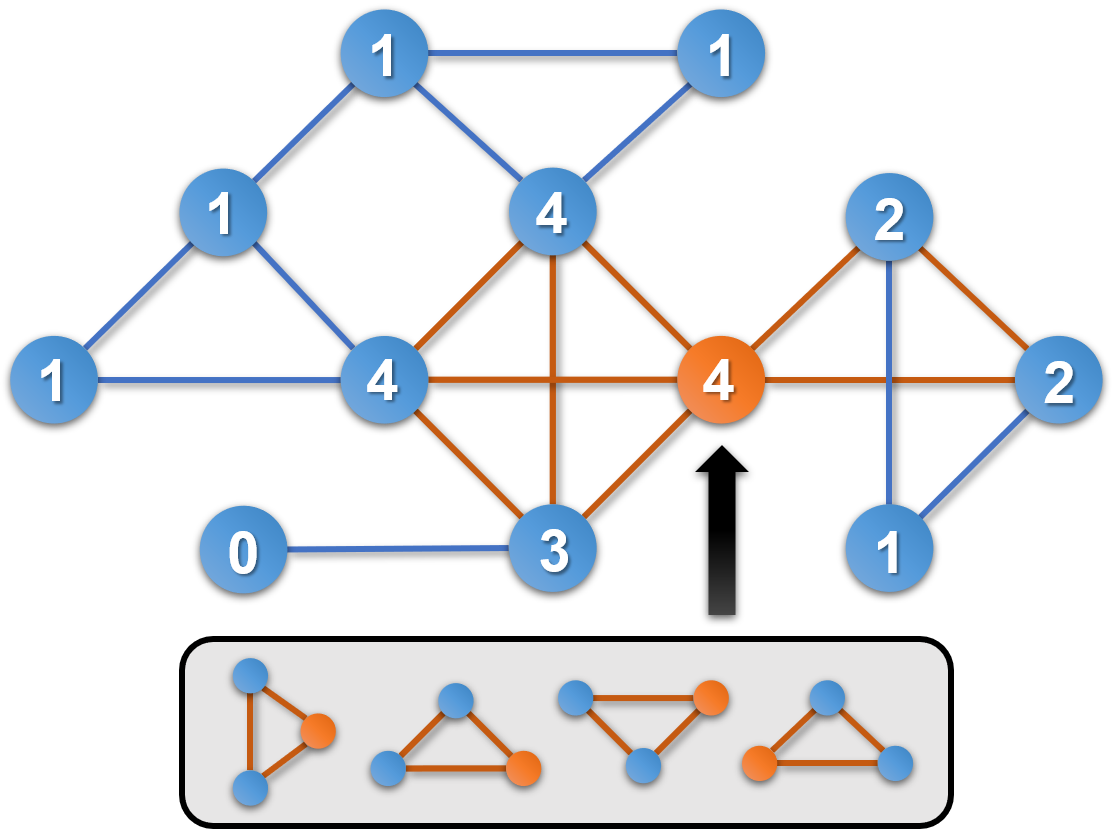}}
	\caption{Schematic diagram of node motif degree in the network. The \textbf{M31-NMD} of the orange node equals to 4, because four triangle motifs \textbf{M31} shown in the box contains this orange node.}
	\label{Fig:showMD}
\end{figure}

In this paper, we utilize the concept of Node Motif Degree \textbf{(NMD)} to characterize the association of a particular multivariate relationship at a node, and its definition is as follows.

\begin{definition}
	\textbf{(Node Motif Degree)} In the graph $G = \{V, E\}$, the node motif degree information $NMD (v)$ of a node $v \in V$ is defined as the number of the motif $M$, whose constituent nodes includes the node $v$.
\end{definition}

As shown in the Figure~\ref{Fig:showMD}, the number of the nodes represents the node motif degree information of the triangle motif (\textbf{M31}). Through these graph features, we can better connect the single node, the network motif and the multivariate relationship in the network structure.

\subsection{The Framework of MORE}


In this part, we will introduce the MORE model proposed in this article, which is simple and easy to implement. This model can effectively catch the multivariate relationship information in the network. While improving the accuracy of graph learning tasks, MORE greatly speeds up the convergence and reduces the time cost in the learning process.

\begin{figure}[!b]
	\centerline{\includegraphics[scale=0.27]{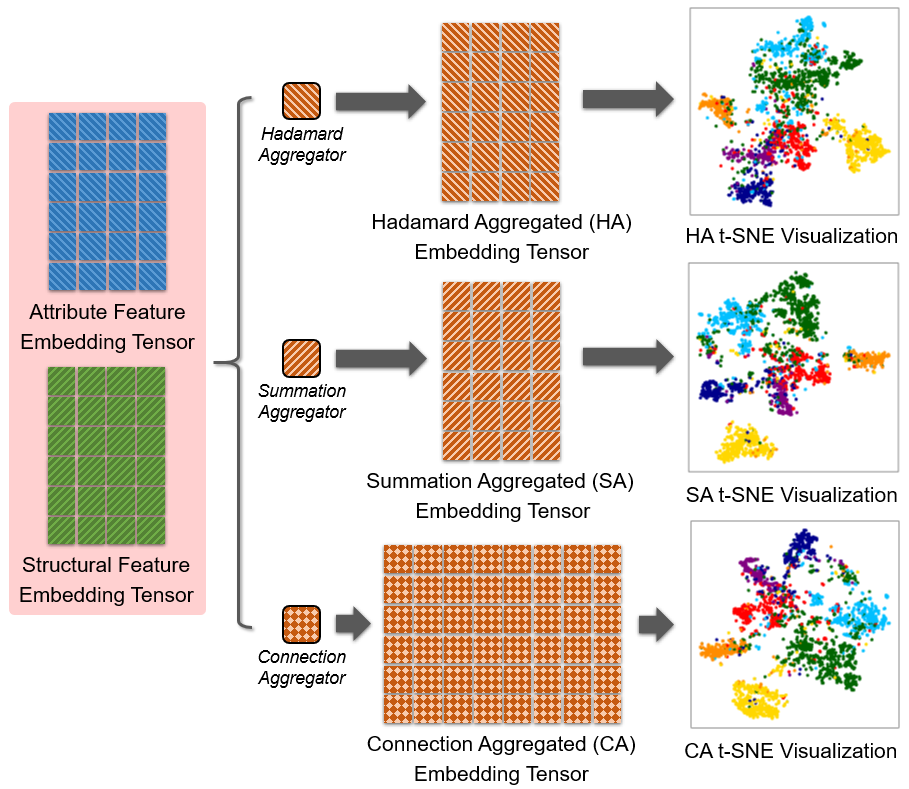}}
	\caption{Schematic diagram of three different multivariate relationship aggregation methods. The t-SNE view on the right shows the embedding results of three different methods on the \textbf{Cora} dataset.}
	\label{Fig:showAgg}
\end{figure}

The framework structure of the entire MORE model is shown in Figure~\ref{Fig:Framework}.  The overall model contains two parts, one is the graph feature information calculation, and the other is the learning and classification process. The attribute feature tensor and structure feature tensor output by the first part will be used as the input of the second part. It is worth noting that in the above figure, due to space constraints, we have not fully expanded the content of the aggregated part (the red dotted box in the figure). The content of the method in this part will be explained in Figure~\ref{Fig:showAgg}.

For the input network dataset, we first extract the characteristic information of its nodes. MORE algorithm considers the attribute features and structural features of the node. The algorithm directly extracts the attribute information of each node in the graph structure, and integrates it into the node Attribute Feature Tensor \textbf{(AFT)} of the total graph. Moreover, to better characterize the structural association and multivariate relationship information, the algorithm calculates five kinds of node motif degree information of each node in the network, including the node motif degree of \textbf{M31}, \textbf{M32}, \textbf{M41}, \textbf{M42} and \textbf{M43}. This information will be integrated with the original degree of the node, and then the node Structural Feature Tensor \textbf{(SFT)} will be obtained.

Because the two feature tensors obtained in the above process often have large scale differences, and the direct connection two tensors will easily destroy the internal correlation of the feature itself, we consider using a graph representation learning model to reduce or expand the dimension of original tensors. By transform its node feature representation method, we can make the two have the same dimensional information. In this paper, this model uses the GCN model, which uses a graph convolutional network to implement the network embedding process. The GCN model is an effective model to deal with graph-related problems. The core of GCN uses a graph convolution operation based on spectral graph theory, and its formula is expressed as follows:
\begin{equation}
	g_{\theta} \times_{G} x = \theta (I_{n} + D^{-\frac {1} {2}} A D^{-\frac {1} {2}})x
\end{equation}
Wherein, $x$ represents the graph feature vector, $\theta$ represents the weight parameter vector, and $\times_{G}$ represents the graph convolution operation. To improve the stability of the algorithm and avoid gradient explosion or disappearance during the learning process, GCN replaced $I_{n} + D^{-\frac {1} {2}} A D^{-\frac {1} {2}}$ in the above formula with $\widetilde{A} = \bar{D}^{-\frac {1} {2}} \bar{A} \bar{D}^{-\frac {1} {2}}$. Wherein, $\bar{A} = I_{n}+A$ and $\bar{D}_{ii} = \sum_j \bar{A}_{ij}$. Through this methodmodel, we transform the original AFT and SFT into their respective embedding tensors, namely the Attribute Feature Embedding Tensor \textbf{(AFET)} and the Structure Feature Embedding Tensor \textbf{(SFET)}.

In order to aggregate the AFET and SFET, get the Aggregated Embedding Tensor \textbf{(AET)}, and apply it to practical tasks, MORE proposed three multivariate relationship aggregation methods, as shown in Figure~\ref{Fig:showAgg}. The Hadamard Aggregation (HA) method, or dot multiplication aggregation method, gets AET by performing dot multiplication of two embedding tensors; The Summation Aggregation (SA) method gets AET by adding two embedding tensors. These two aggregation methods will not change the size of the aggregation tensor, consume less time for the iterative process, and will not greatly affect the internal correlation information of the features. The Connection Aggregation (CA) method gets AET by connecting two embedding tensors. Although this method will double the size of the aggregation tensor, and the iterative process requires a higher time cost, it can completely retain all the feature information in the total network dataset. In this paper, we use \textbf{MORE-HA}, \textbf{MORE-SU}, and \textbf{MORE-CO}, to represent the MORE algorithm using three aggregation methods (HA, SA and CA) in the aggregation processing of AFET and SFET, respectively.

Finally, the MORE model is used for node classification tasks to measure and express the performance of our algorithm in real network environments. We have considered inputting AET into a graph learning model again for iterative training. However, it was found through experiments that, when the GCN model was applied, the previous iterative process was sufficient to meet the needs of the processing task. If we add more graph learning models, it will cause severe overfitting. To this end, the algorithm directly adds a Softmax layer at the end to generate the one-hot vector required for node classification.

The above is an iterative process of the MORE model. This process can be expressed by the following formula.
\begin{equation}
\hat{Y} = SM ( \widetilde{A} \cdot Aggre ( GCN(AFT), GCN(SFT)) \cdot  \Theta )
\end{equation}
Among them, $GCN(\cdot)$ represents the single GCN iteration process, and $Aggre(\cdot)$ represents the aggregation process, which is selected from three methods. $SM(\cdot)$ represents the Softmax function, which is defined as $SM(x_{i})=\exp{(x_i)} / \sum_{i} \exp{(x_i)}$. Next, we use the cross-entropy error as the loss function of the model in the node classification task, and continuously adjust the values of the relevant weight parameters through the gradient descent algorithm. The specific error function of one data $\mathcal{L}_{i}$ can be described by the following formula.
\begin{equation}
\mathcal{L}_{i} = -  \sum_{l=1}^{\#label}Y_{il}\cdot \ln(\hat{Y}_{il})
\end{equation}
Wherein, $Y$ represents the true node classification label of the network data set, and $\hat{Y}$ represents the node classification label predicted by the model.

\section{Experiment}
\label{Section:4}

In this section, we explain the experimental design of the node classification task using the MORE model. We illustrate the data set used firstly. Then, the operating environment, hyperparameter settings, and baseline method will be introduced. Finally, the advantages of our method are explained in terms of the accuracy and efficiency of the node classification task.

\subsection{Datasets}

We selected 6 different datasets to conduct experiments to reflect the characteristics and advantages of our model. The basic information of these data sets is shown in Table~\ref{Table:DS}. The information of each dataset is divided into three parts: the first part is the basic information, the second part is the structure information, and the third part is the network motif information. In this tabel, \textbf{\#Node} represents the number of nodes in the dataset network, \textbf{\#Edge} represents the number of edges, \textbf{\#Feature} represents the feature number of each node, and \textbf{\#Label} represents the type number of node labels in the total network. `-' in \textbf{\#Feature} means that the dataset is missing node feature data. In addition to the above basic network information, we have additionally listed the average degree $\bm{\bar{d}}$, the maximum degree $\bm{\max{(d)}}$, the network density $\bm{\rho}$, and the overall clustering coefficient \textbf{Clustering} of the network. These statistical indicators will effectively characterize the density of associations between data and the connection status of the network.

Because our model needs to consider the number of motifs in the network, and use this to characterize the multivariate relationships and high-order structural features in the dataset, we need to count several representative network motifs and get the motif node degree information of each node in the network. The number of network motifs we used in each network dataset is also shown in Table~\ref{Table:DS}. \textbf{\#M31} represents the number of triangle motif, and \textbf{\#M32} represents the number of three-order path motif in the network. \textbf{\#M41}, \textbf{\#M42} and \textbf{\#M43} respectively represent the number of three kinds of four-order motif mentioned in this paper.

\begin{table}[!t]
	\centering
	\caption{Datasets used in the experiment.}
	\label{Table:DS}
	\resizebox{0.44\textwidth}{!}{
		\begin{threeparttable}{
		\begin{tabular}{cccc}
			\toprule
			\textbf{Network Name} & \textbf{Cora$\bm{^1}$}     & \textbf{Email-Eucore$\bm{^3}$} & \textbf{Facebook$\bm{^4}$}     \\
			\midrule
			\textbf{\#Node}       & 2708              & 1005                  & 6637                  \\
			\textbf{\#Edge}       & 5278              & 16706                 & 249967                \\
			\textbf{\#Feature}    & 1433              & -                     & -                     \\
			\textbf{\#Label}      & 7                 & 41                    & 3                     \\
			\midrule
			$\bm{\bar{d}}$           & 3.9               & 33.25                 & 75.32                 \\
			$\bm{\max{(d)}}$         & 168               & 347                   & 840                   \\
			$\bm{\rho}$            & 0.00144           & 0.033113              & 0.011351              \\
			\textbf{Clustering}   & 0.24              & 0.399                 & 0.278                 \\
			\midrule
			\textbf{\#M31}   & 1630              & 105461                & 2310053               \\
			\textbf{\#M32}       & 47411             & 866833                & 30478345              \\
			\textbf{\#M41}        & 220               & 423750                & 13743319              \\
			\textbf{\#M42}        & 2468              & 2470220               & 83926778              \\
			\textbf{\#M43}        & 1536              & 909289                & 45518420              \\
			\bottomrule
			\toprule
			\textbf{Network Name} & \textbf{Polblogs$\bm{^2}$} & \textbf{Football$\bm{^2}$}     & \textbf{TerrorAttack$\bm{^1}$} \\
			\midrule
			\textbf{\#Node}       & 1224              & 115                   & 1293                  \\
			\textbf{\#Edge}       & 16718             & 613                   & 3172                  \\
			\textbf{\#Feature}    & -                 & -                     & 106                   \\
			\textbf{\#Label}      & 2                 & 12                    & 6                     \\
			\midrule
			$\bm{\bar{d}}$            & 27.32             & 10.66                 & 4.91                  \\
			$\bm{\max{(d)}}$         & 351               & 12                    & 49                    \\
			$\bm{\rho}$             & 0.022336          & 0.093516              & 0.003798              \\
			\textbf{Clustering}   & 0.32              & 0.403                 & 0.378                 \\
			\midrule
			\textbf{\#M31}   & 101043            & 810                   & 26171                 \\
			\textbf{\#M32}       & 1038396           & 3537                  & 232                   \\
			\textbf{\#M41}        & 422327            & 732                   & 241419                \\
			\textbf{\#M42}        & 2775480           & 1155                  & 0                     \\
			\textbf{\#M43}        & 1128796           & 564                   & 0                    \\
			\bottomrule
	\end{tabular}}
	\begin{tablenotes}
		\footnotesize
		\item \textbf{1} \url{https://linqs.soe.ucsc.edu/node/236}
		\item \textbf{2} \url{http://www-personal.umich.edu/~mejn/netdata/}
		\item \textbf{3} \url{http://snap.stanford.edu/data/email-Eu-core.html}
		\item \textbf{4} \url{http://networkrepository.com/fb-CMU-Carnegie49.php}
	\end{tablenotes}
	\end{threeparttable}}
\end{table}

\subsection{Environmental Settings}

Because the comparison of the running time is needed in this experiment, we provide information for the computer's operating environment and hardware configuration. We use a Hewlett-Packard PC, ENVY-13: The operating system is Windows 10 64-bit, the CPU is Intel i5-8265U, and the memory is 8G. We use Python 3.7.3 and Tensorflow 1.14.0 for coding, and use Visual Studio Code for programming.

During the training process, we divide the original dataset into a training set, a validation set, and a test set. If the number of nodes in the dataset is between 1150 and 3000, we randomly select \textbf{150} data as the training set (\textbf{TRS}) and \textbf{500} data as the verification set (\textbf{VAS}), and finally test the model effect on a test set (\textbf{TES}) consisting of \textbf{500} data. For the \textbf{Email-Eucore}, \textbf{Facebook}, and \textbf{Football}, we construct three sets at similar proportions. Besides, since the attribute information of the nodes is used in the iteration, we use a method that assigns a one-hot vector to each node randomly for the network without features. This method is simple, easy to implement, and almost does not affect overall mission performance.

The choice of hyperparameters is also worth considering because they directly affect the final experimental results. In this experiment, if there is no special statement, we optimize the model using \textbf{Adam} with the learning rate $\bm{\alpha}$ of \textbf{0.01}, and the maximum training iterations number \textbf{max\_epoch} of \textbf{300}. This experiment uses two regularization methods: random inactivation and L2 regularization. The probability of random inactivation \textbf{drop\_out} is \textbf{0.5}, and the L2 regularization factor $\bm{\lambda}$ is \textbf{0.0005}. To avoid the occurrence of overfitting, the early stopping method is adopted, and the tolerance is \textbf{30} (If the loss of the validation set does not decrease during 30 consecutive iterations, the algorithm will automatically stop).  Also, the embedding dimension \textbf{ED} is an important attribute, which directly determines the capacity of the model space. In this experiment, the graph representation learning model use \textbf{one layer} of GCN structure. AFT and SFT are embedded in \textbf{32}, \textbf{64}, \textbf{128}, \textbf{256}, and \textbf{512} dimensions, respectively, and the best performance is selected as the final experimental result.

\begin{table*}[!t]
	\centering
	\caption{GCN, MORE-HA, MORE-SU and MORE-CO model accuracy of node classification tasks on 6 network datasets. The bolded data indicate the best performing method in the set of parameters on the network.}
	\label{Table:RA}
	\resizebox{0.98\textwidth}{!}{
		\begin{tabular}{ccccc|cccc|cccc}
			\toprule
			\multirow{2}{*}{\textbf{Network}} & \multicolumn{4}{c}{$\bm{\alpha=0.01, max\_epoch= 300}$}                          & \multicolumn{4}{c}{$\bm{\alpha=0.001, max\_epoch= 500}$}                 & \multicolumn{4}{c}{$\bm{\alpha=0.0003, max\_epoch= 1000}$}              \\ \cmidrule(lr){2-5} \cmidrule(lr){6-9} \cmidrule(lr){10-13}
			& \multicolumn{1}{|c}{\textbf{GCN}} & \textbf{MORE-HA} & \textbf{MORE-SU} & \textbf{MORE-CO} & \textbf{GCN} & \textbf{MORE-HA} & \textbf{MORE-SU} & \textbf{MORE-CO} & \textbf{GCN} & \textbf{MORE-HA} & \textbf{MORE-SU} & \textbf{MORE-CO} \\
			\midrule
			\multicolumn{1}{c|}{\textbf{Cora}}                                                                    & \textbf{82.20\%}      & 81.80\%     & 81.40\%     & 81.60\%     & 79.00\%      & \textbf{82.30\%}     & 80.40\%     & 81.60\%     & 76.20\%      & 81.20\%     & 79.90\%     & \textbf{81.40\% }    \\
			\multicolumn{1}{c|}{\textbf{Email-Eucore}}                                                            & 51.50\%      & \textbf{61.50\%}    & 58.25\%     & 60.75\%     & 23.00\%      & 53.50\%     & 54.00\%     & \textbf{56.00\%}     & 09.75\%       & 47.00\%     & 50.25\%     & \textbf{51.75\% }    \\
			\multicolumn{1}{c|}{\textbf{Facebook}}                                                                & 58.20\%      & 60.05\%     & \textbf{61.45\%}     & 60.40\%     & 53.70\%      & \textbf{58.65\% }    & 56.65\%     & 54.65\%     & 53.85\%      & \textbf{55.70\%}     & 54.20\%     & 54.65\%     \\
			\multicolumn{1}{c|}{\textbf{Polblogs}}                                                                & 95.80\%      & \textbf{96.20\%}     & 95.80\%     & 95.90\%     & 96.00\%      & \textbf{96.20\%}     & 95.80\%     & 95.80\%     & 95.80\%      & \textbf{96.20\% }    & 95.80\%     & 95.60\%     \\
			\multicolumn{1}{c|}{\textbf{Football}}                                                                & \textbf{86.67\%}      & \textbf{86.67\%}     & \textbf{86.67\%}     &\textbf{ 86.67\% }    & 44.44\%      & \textbf{86.67\%}     & \textbf{86.67\% }    & \textbf{86.67\%}     & 40.00\%      & \textbf{86.67\% }    & 84.44\%     &\textbf{ 86.67\%}     \\
			\multicolumn{1}{c|}{\textbf{TerrorAttack}}                                                            & 75.20\%      & 35.00\%     & 76.20\%     & \textbf{76.40\%}     & 71.20\%      & 33.80\%     & 75.00\%     & \textbf{76.20\%}     & 66.40\%      & 33.60\%     & 75.20\%     & \textbf{76.40\%}    \\
			\bottomrule
	\end{tabular}}
\end{table*}

\begin{table*}[!t]
	\centering
	\caption{Efficiency comparison of GCN, MORE-HA, MORE-SU and MORE-CO model in the case of convergence in 6 network datasets. The hyperparameters are set to $\bm{\alpha}$ \textbf{= 0.003}, \textbf{ED = 256} (to guarantee horizontal comparison can be carried out), and \textbf{max\_epoch = 2000} (to ensure that the algorithm runs to local convergence).}
	\label{Table:RE}
	\resizebox{0.98\textwidth}{!}{
		\begin{tabular}{c|cccccc|c|cccccc}
			\toprule
			\textbf{Network}                       & \textbf{Model} & \textbf{Accuracy} & \textbf{\#Iter} & \textbf{ASTT(s)} & \textbf{OIT(s)} & \textbf{TET(s)} & \textbf{Network}                       & \textbf{Model} & \textbf{Accuracy} & \textbf{\#Iter} & \textbf{ASTT(s)} & \textbf{OIT(s)} & \textbf{TET(s)} \\
			\midrule
			\multirow{4}{*}{\textbf{Cora}}         & \textbf{GCN}            & 82.10\%           & 791                   & 0.0622                                                                             & 71.8276                                                                    & 0.0301                     & \multirow{4}{*}{\textbf{Polblogs}}     & \textbf{GCN}            & 95.60\%           & 1008                  & 0.0458                                                                             & 62.2642                                                                    & 0.0175                     \\
			
			& \textbf{MORE-HA}             & 81.30\%           & 338                   & 0.0887                                                                             & 42.9283                                                                    & 0.0322                     &                                        & \textbf{MORE-HA}             & 96.20\%           & 315                   & 0.0586                                                                             & 25.4139                                                                    & 0.0309                     \\
			& \textbf{MORE-SU}             & 79.90\%           & 251                   & 0.0867                                                                             & 31.8249                                                                    & 0.0359                     &                                        & \textbf{MORE-SU}             & 95.80\%           & 124                   & 0.0732                                                                             & 12.5456                                                                    & 0.0346                     \\
			& \textbf{MORE-CO}             & 81.10\%           & 257                   & 0.1047                                                                             & 39.1086                                                                    & 0.0583                     &                                        & \textbf{MORE-CO}             & 95.80\%           & 157                   & 0.0812                                                                             & 17.7403                                                                    & 0.0319                     \\
			\midrule
			\multirow{4}{*}{\textbf{Email-Eucore}} & \textbf{GCN}            & 51.00\%           & 954                   & 0.0434                                                                             & 56.0472                                                                    & 0.0170                     & \multirow{4}{*}{\textbf{Football}}     & \textbf{GCN }           & 84.44\%           & 559                   & 0.0142                                                                             & 9.9511                                                                     & 0.0140                     \\
			& \textbf{MORE-HA}             & 59.25\%           & 350                   & 0.0417                                                                             & 20.1162                                                                    & 0.0160                     &                                        & \textbf{MORE-HA}             & 86.67\%           & 164                   & 0.0149                                                                             & 3.1719                                                                     & 0.0123                     \\
			& \textbf{MORE-SU}             & 54.75\%           & 220                   & 0.0376                                                                             & 11.8910                                                                    & 0.0160                     &                                        & \textbf{MORE-SU}             & 84.44\%           & 138                   & 0.0145                                                                             & 2.5635                                                                     & 0.0044                     \\
			& \textbf{MORE-CO}             & 57.25\%           & 220                   & 0.0507                                                                             & 17.0667                                                                    & 0.0259                     &                                        & \textbf{MORE-CO}             & 86.67\%           & 99                    & 0.0171                                                                             & 2.2097                                                                     & 0.1577                     \\
			\midrule
			\multirow{4}{*}{\textbf{Facebook}}     & \textbf{GCN}           & 57.60\%           & 495                   & 0.5414                                                                             & 360.3334                                                                   & 0.1871                     & \multirow{4}{*}{\textbf{TerrorAttack}} & \textbf{GCN}            & 75.40\%           & 327                   & 0.0236                                                                             & 10.9818                                                                    & 0.0110                     \\
			& \textbf{MORE-HA}             & 59.05\%           & 167                   & 0.5475                                                                             & 123.9314                                                                   & 0.2079                     &                                        & \textbf{MORE-HA}             & 33.80\%           & 146                   & 0.0263                                                                             & 5.3676                                                                     & 0.0110                     \\
			& \textbf{MORE-SU}             & 59.55\%           & 102                   & 0.5890                                                                             & 81.4163                                                                    & 0.1985                     &                                        & \textbf{MORE-SU}             & 76.20\%           & 86                    & 0.0374                                                                             & 4.4990                                                                     & 0.0150                     \\
			& \textbf{MORE-CO}             & 59.80\%           & 82                    & 0.6361                                                                             & 70.6593                                                                    & 0.2224                     &                                        & \textbf{MORE-CO}             & 76.20\%           & 87                    & 0.0333                                                                             & 4.0502                                                                     & 0.0140                    \\
			\bottomrule
	\end{tabular}}
\end{table*}

\subsection{Baseline Method}

We compare the MORE model with the GCN model proposed by Kipf et al~\cite{Kipf2016Semi-Supervised}. As a well-known algorithm in the field of graph convolutional networks, GCN is based on spectral graph theory and has been recognized by many researchers in this field. We think that using it as a comparison algorithm can effectively illustrate the advantages of our algorithm. During the experimental phase, the GCN model used for comparison adopts the double-layer structure as the original paper. The specific iteration formula is as follows:
\begin{equation}
\hat{Y} = SM ( \widetilde{A} \cdot ReLU(\widetilde{A}XW^{(0)})\cdot W^{(1)})
\end{equation}
Wherein, \textbf{ReLU} represents the linear rectification function which used as the activation function, and $\bm{W^{(0)}}$ and $\bm{W^{(1)}}$ represent the training weight matrix.

\subsection{Node Classification Accuracy}

On the whole, we will show our experimental content in terms of the accuracy and overall efficiency of the node classification task. For experiments on the accuracy, we set three groups of hyperparameters that differ in the learning rate $\bm{\alpha}$ and the maximum number of iterations \textbf{max\_epoch} (Their values are [0.01, 300], [0.001, 500], and [0.0003, 1000], respectively), to enhance the persuasiveness of our experimental results. Then, we iterated the MORE model and GCN model using three groups of hyperparameters in six network datasets, and the experimental results are shown in Table~\ref{Table:RA}. It can be seen that in all networks, the highest accuracy of the MORE model is better than that of the GCN model. In the \textbf{Email-Eucore} dataset which is rich in the network motif, our model has improved by \textbf{10\%} based on the 51.50\% accuracy of the GCN model.

\begin{figure*}[!t]
	\begin{minipage}{0.32\textwidth}
		\centerline{\includegraphics[width=6.0cm]{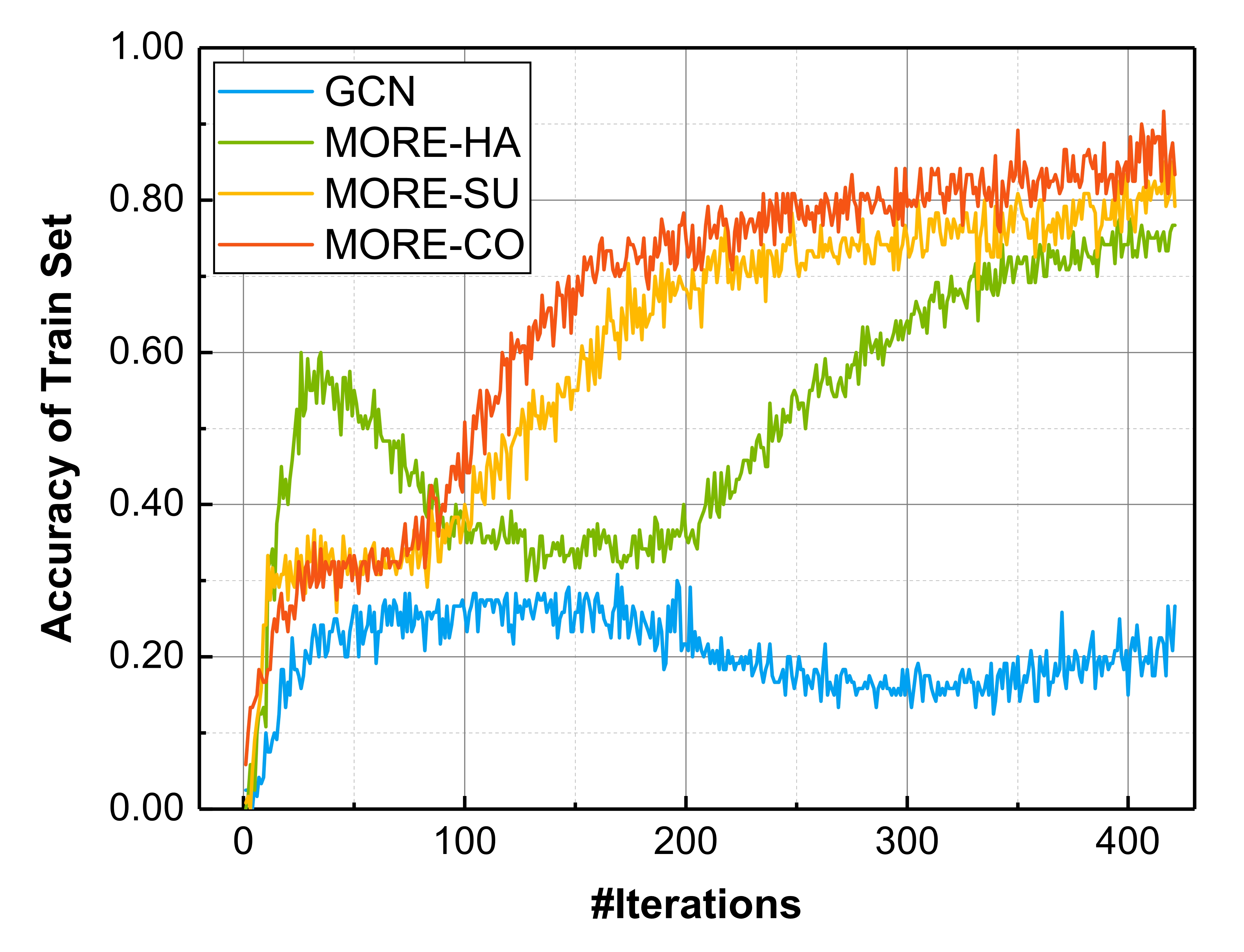}}
		\centering{(a) Change in Accuracy on TRS.}
	\end{minipage}
	\hfill
	\begin{minipage}{0.32\textwidth}
		\centerline{\includegraphics[width=6.0cm]{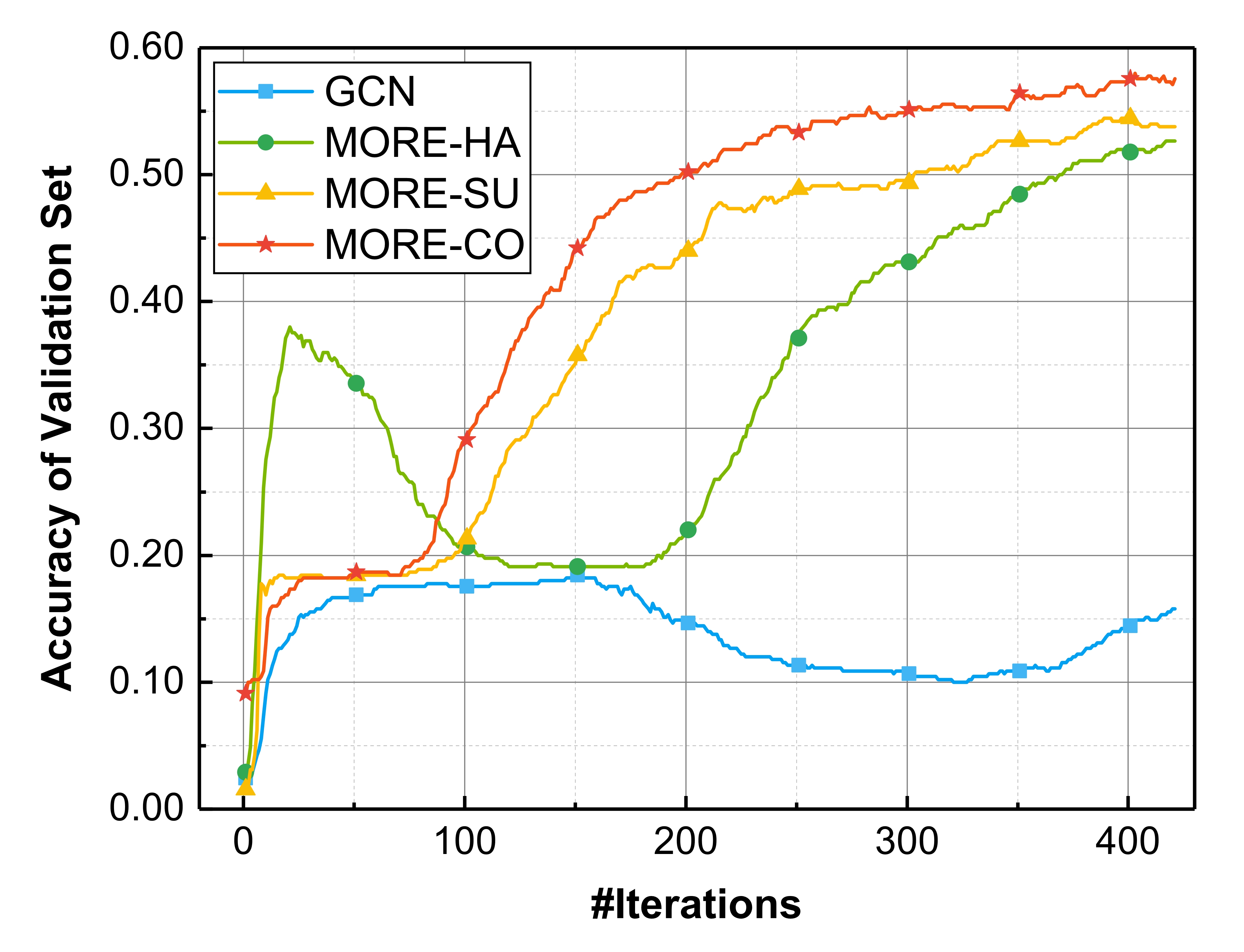}}
		\centering{(b) Change in Accuracy on VAS.}
	\end{minipage}
	\hfill
	\begin{minipage}{0.32\textwidth}
		\centerline{\includegraphics[width=6.0cm]{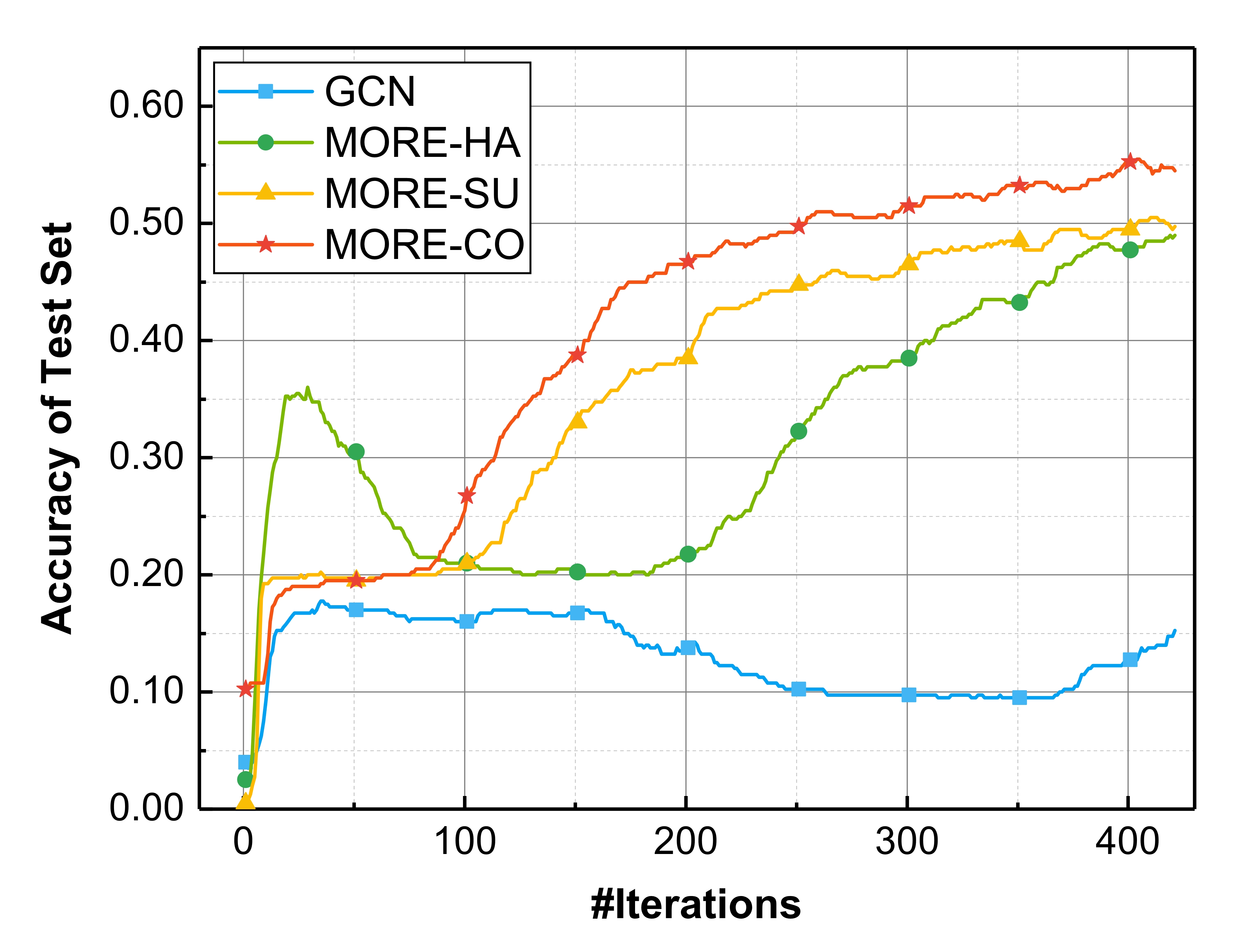}}
		\centering{(c) Change in Accuracy on TES.}
	\end{minipage}
	\vfill
	\begin{minipage}{0.32\textwidth}
		\centerline{\includegraphics[width=6.0cm]{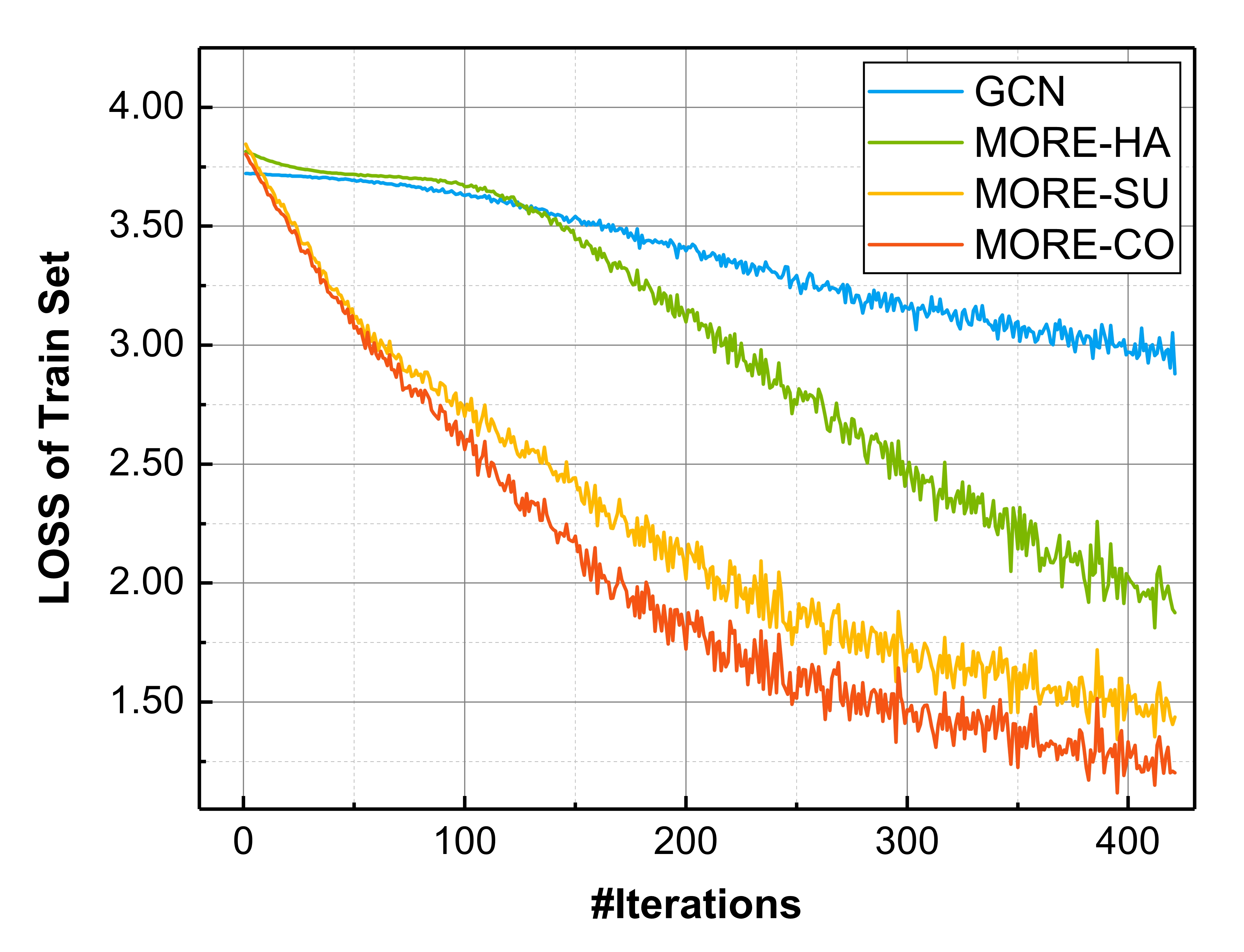}}
		\centering{(d) Change in LOSS on TRS.}
	\end{minipage}
	\hfill
	\begin{minipage}{0.32\textwidth}
		\centerline{\includegraphics[width=6.0cm]{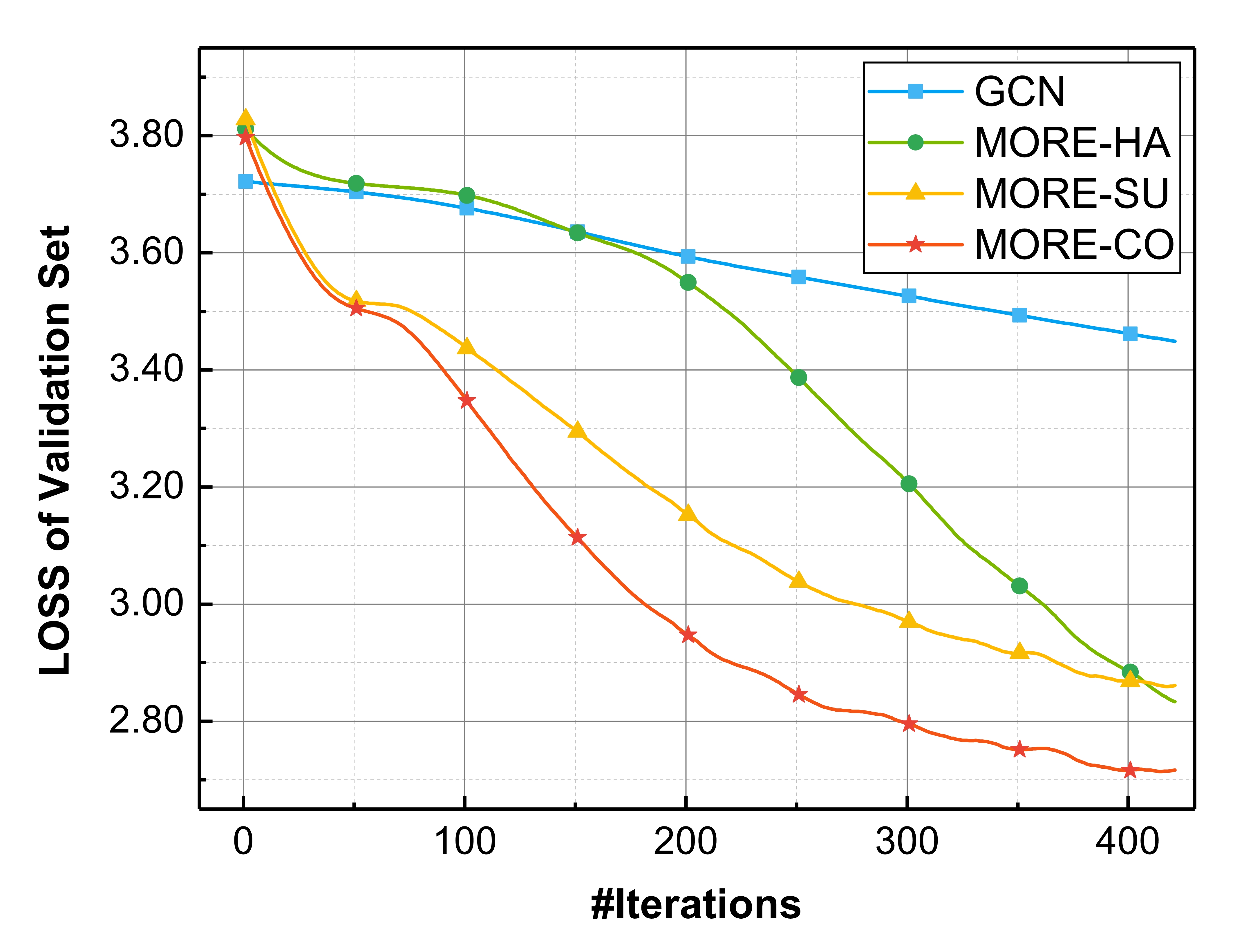}}
		\centering{(e) Change in LOSS on VAS.}
	\end{minipage}
	\hfill
	\begin{minipage}{0.32\textwidth}
		\centerline{\includegraphics[width=6.0cm]{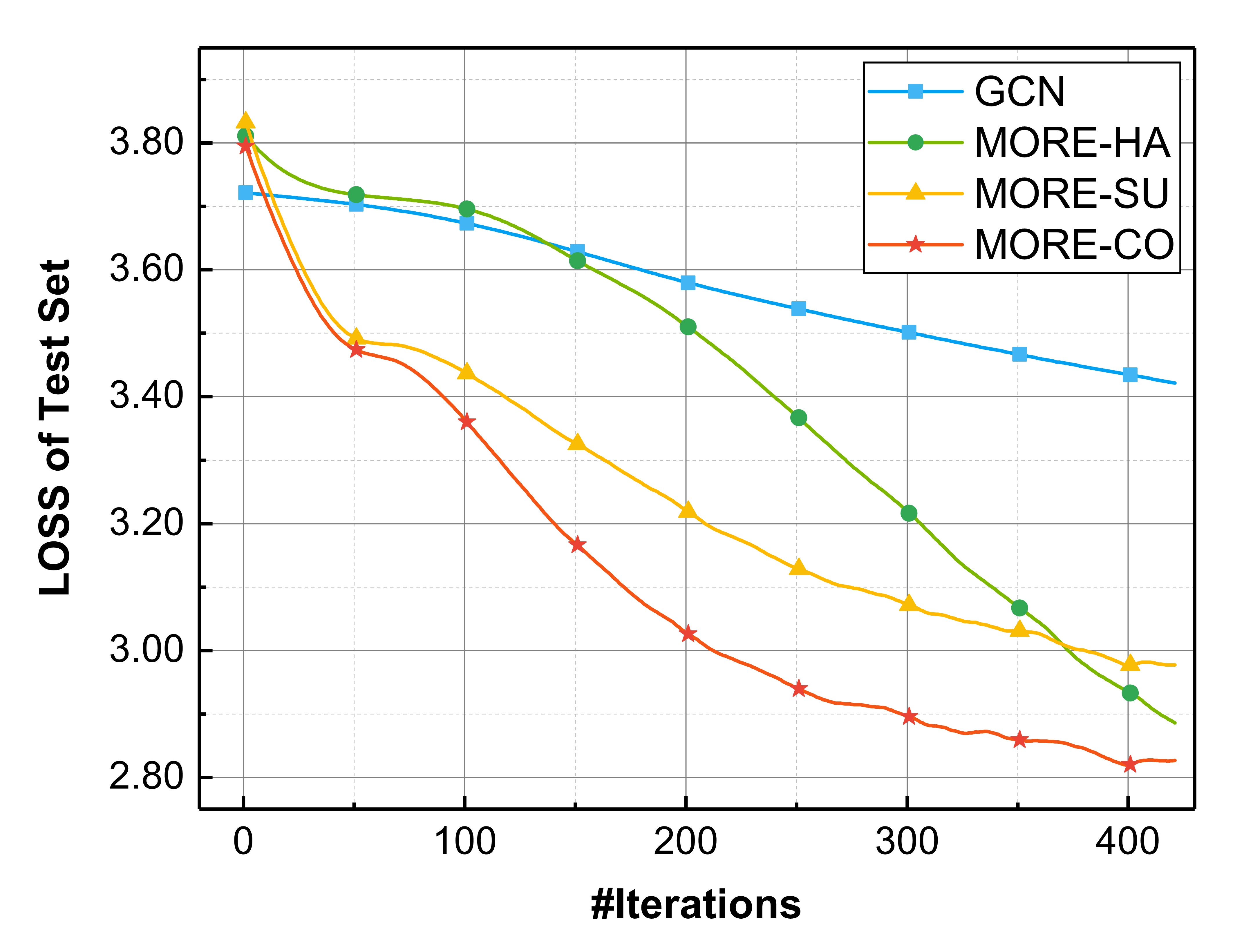}}
		\centering{(f) Change in LOSS on TES.}
	\end{minipage}
	\caption{Comparison diagram of the convergence speed of the four models on \textbf{Email-Eucore} dataset. The hyperparameters of the running environment are set to $\bm{\alpha}$ \textbf{= 0.001}, \textbf{max\_epoch = 500}, and \textbf{ED = 256}.}
	\label{Fig:showIE}
\end{figure*}

By comparing the three models proposed in this paper, namely MORE-HA, MORE-SU, and MORE-CO, we can obviously find that the MORE-HA model has the highest times of getting the best accuracy. This is because the MORE-HA model uses the Hadamard product process to make one type of feature information as a bias of another type, and preserves the implicit association as much as possible. The aggregation process of the MORE-SU model and the MORE-CO model leads to the cancellation or even destruction of some of this correlation information, resulting in a decrease in the overall accuracy. However, it is worth noting that in \textbf{TerrorAttack} dataset, we find that the accuracy of the MORE-HA  is very low. The reason for this phenomenon is that the number of \textbf{M42} and \textbf{M43} motif in the network is \textbf{zero}, causing the Hadamard product to completely lose data. This phenomenon does not occur during the summation process and connection process, so it can be said that the MORE-SU and MORE-CO models have stronger stability than MORE-HA model.

\subsection{Node Classification Effectiveness}

We perform comparative experiments on the efficiency of the node classification task. We iterate the four models to convergence with a learning rate $\bm{\alpha}$ of \textbf{0.003}. The experimental results are shown in Table~\ref{Table:RE}. Wherein, we use the number of iterations \textbf{\#Iter}, the Average Single Training Time (\textbf{ASTT}), the Overall Iteration Time (\textbf{OIT}), and the TEst running Time (\textbf{TET}) to measure the convergence speed and training efficiency of the model. It can be found that by combining the multivariate relationships in the network environment, the overall efficiency of the algorithm has been greatly improved. Compared with the GCN model, the MORE model can reduce the \textbf{OIT} required in the iterative process by up to \textbf{19.5\%} of GCN, and reduce \textbf{\#Iter} by up to \textbf{15.5\%} of GCN. Moreover, the MORE model has not reduced the accuracy or even improved it in the node classification tasks of most datasets.

Figure~\ref{Fig:showIE} further shows the convergence speed comparison between MORE and GCN models. We run four models on the \textbf{Email-Eucore} dataset with $\bm{\alpha}$ \textbf{= 0.001}, and continue to iterate until one of the models converges. In this process, we calculate the accuracy and loss of the four models on the training set, validation set, and test set for node classification tasks after each iteration, and draw it into a line chart. It can be clearly seen that the convergence speed of the MORE model is significantly higher than that of the GCN model. And compared with MORE-HA and MORE-SU, because the node information is completely preserved during the connection process, MORE-CO using the connection aggregation has more information available during the iteration. This reason ultimately leads that MORE-CO has the highest training efficiency.

It is worth noting that we can clearly find from Figure~\ref{Fig:showIE} (a)-(c) that around the 25th iteration, the accuracy trends of all models have shown a large degree of smoothness. Even in the MORE-HA model, the accuracy of the node classification task experienced a local peak. This phenomenon is caused by the data set and the gradient descent algorithm. In the \textbf{Email-Eucore} dataset, comparing the four models, it can be found that MORE-SU and MORE-CO are less affected. They did not experience a sudden increase or decrease in the accuracy of the validation set and the test set. This matches the conclusions we have drawn above, that is, MORE-SU and MORE-CO are more stable than MORE-HA.

\section{Discussion}
\label{Section:5}

In this section, we discuss the future direction of improvement about the current model. For our model, we propose three feasible directions to meet the needs of more data processing tasks.

\begin{itemize}

 \item \textbf{Use other graph representation learning model.} Our method has the potential as a framework. The MORE model proposed in this paper uses a single-layer GCN structure as an intermediate module for graph representation learning and only compares it with the GCN model. Next, we will try to improve MORE into a framework that can be applied to graph learning algorithms to improve the comprehensive performance of GCNN models.
	
 \item \textbf{Extend the application scope of the graph structure.} Due to the limitation of the GCN algorithm, the currently proposed MORE model can only be applied to undirected graph structures. In the future, we hope to extend existing models to more graph structures, such as directed graphs. There are many types of directed network motif and they can more specifically represent some kinds of multivariate relationships in the network. We believe that it can further improve the accuracy and efficiency of the method.
	
 \item \textbf{Apply to large-scale networks.} The existing graph learning methods all face the challenge that it is difficult to apply to large-scale networks. This is due to multiple reasons such as low iteration efficiency, high memory consumption, and poor parallelism in a large-scale network environment. With the continuous increase of data, graph learning of large-scale networks is an inevitable development trend in the future. We want to improve the applicability of the algorithm in large-scale networks to meet the needs of an ever-increasing network dataset in the future work.

\end{itemize}

\section{Conclusion}
\label{Section:6}

Multivariate relationships appear frequently in various types of real networks. In this paper, we propose a new node classification method based on graph structure and multivariate relations. Our MORE model utilizes the network motif to characterize multivariate relations in the network environment. This relation information that is partially missed or even completely lost in other graph learning algorithms is effectively used in MORE, which makes our algorithm more accurate and efficient. MORE transforms the multivariate relation into the node structural characteristics, and aggregates with the node attribute characteristics, and completes the task of predicting the types of nodes. Experimental results on several network datasets show that the MORE model can effectively capture the potential multivariate relationship information. In this setting, our model is comprehensively superior to the GCN method in terms of computational efficiency and accuracy. In future work, we will try to expand MORE to a graph learning framework, and improve the adaptability of various graph structures and large-scale networks.

\clearpage

\bibliographystyle{ACM-Reference-Format}
\bibliography{References}


\begin{thebibliography}{45}


\ifx \showCODEN    \undefined \def \showCODEN     #1{\unskip}     \fi
\ifx \showDOI      \undefined \def \showDOI       #1{#1}\fi
\ifx \showISBNx    \undefined \def \showISBNx     #1{\unskip}     \fi
\ifx \showISBNxiii \undefined \def \showISBNxiii  #1{\unskip}     \fi
\ifx \showISSN     \undefined \def \showISSN      #1{\unskip}     \fi
\ifx \showLCCN     \undefined \def \showLCCN      #1{\unskip}     \fi
\ifx \shownote     \undefined \def \shownote      #1{#1}          \fi
\ifx \showarticletitle \undefined \def \showarticletitle #1{#1}   \fi
\ifx \showURL      \undefined \def \showURL       {\relax}        \fi
\providecommand\bibfield[2]{#2}
\providecommand\bibinfo[2]{#2}
\providecommand\natexlab[1]{#1}
\providecommand\showeprint[2][]{arXiv:#2}

\bibitem[\protect\citeauthoryear{Bruna, Zaremba, Szlam, and Lecun}{Bruna
  et~al\mbox{.}}{2014}]%
        {bruna2014spectral}
\bibfield{author}{\bibinfo{person}{Joan Bruna}, \bibinfo{person}{Wojciech
  Zaremba}, \bibinfo{person}{Arthur Szlam}, {and} \bibinfo{person}{Yann
  Lecun}.} \bibinfo{year}{2014}\natexlab{}.
\newblock \showarticletitle{Spectral networks and locally connected networks on
  graphs}. In \bibinfo{booktitle}{\emph{International Conference on Learning
  Representations (ICLR2014), CBLS, April 2014}}.
\newblock


\bibitem[\protect\citeauthoryear{Caigny, Coussement, and Bock}{Caigny
  et~al\mbox{.}}{2018}]%
        {Caigny2018new}
\bibfield{author}{\bibinfo{person}{Arno~De Caigny}, \bibinfo{person}{Kristof
  Coussement}, {and} \bibinfo{person}{Koen W.~De Bock}.}
  \bibinfo{year}{2018}\natexlab{}.
\newblock \showarticletitle{A new hybrid classification algorithm for customer
  churn prediction based on logistic regression and decision trees}.
\newblock \bibinfo{journal}{\emph{European Journal of Operational Research}}
  \bibinfo{volume}{269}, \bibinfo{number}{2} (\bibinfo{year}{2018}),
  \bibinfo{pages}{760 -- 772}.
\newblock
\showISSN{0377-2217}


\bibitem[\protect\citeauthoryear{Chang, Saha, Castro-Lacouture, and Yang}{Chang
  et~al\mbox{.}}{2019}]%
        {chang2019Multivariat}
\bibfield{author}{\bibinfo{person}{Soowon Chang}, \bibinfo{person}{Nirvik
  Saha}, \bibinfo{person}{Daniel Castro-Lacouture}, {and}
  \bibinfo{person}{Perry Pei-Ju Yang}.} \bibinfo{year}{2019}\natexlab{}.
\newblock \showarticletitle{Multivariate relationships between campus design
  parameters and energy performance using reinforcement learning and parametric
  modeling}.
\newblock \bibinfo{journal}{\emph{Applied Energy}}  \bibinfo{volume}{249}
  (\bibinfo{year}{2019}), \bibinfo{pages}{253 -- 264}.
\newblock
\showISSN{0306-2619}


\bibitem[\protect\citeauthoryear{Cho, van Merrienboer, G{\"{u}}l{\c{c}}ehre,
  Bahdanau, Bougares, Schwenk, and Bengio}{Cho et~al\mbox{.}}{2014}]%
        {Cho2014GRU}
\bibfield{author}{\bibinfo{person}{Kyunghyun Cho}, \bibinfo{person}{Bart van
  Merrienboer}, \bibinfo{person}{{\c{C}}aglar G{\"{u}}l{\c{c}}ehre},
  \bibinfo{person}{Dzmitry Bahdanau}, \bibinfo{person}{Fethi Bougares},
  \bibinfo{person}{Holger Schwenk}, {and} \bibinfo{person}{Yoshua Bengio}.}
  \bibinfo{year}{2014}\natexlab{}.
\newblock \showarticletitle{Learning Phrase Representations using {RNN}
  Encoder-Decoder for Statistical Machine Translation}. In
  \bibinfo{booktitle}{\emph{Proceedings of the 2014 Conference on Empirical
  Methods in Natural Language Processing, {EMNLP} 2014, October 25-29, 2014,
  Doha, Qatar, {A} meeting of SIGDAT, a Special Interest Group of the {ACL}}}.
  \bibinfo{pages}{1724--1734}.
\newblock


\bibitem[\protect\citeauthoryear{Chorowski, Bahdanau, Serdyuk, Cho, and
  Bengio}{Chorowski et~al\mbox{.}}{2015}]%
        {Chorowski2015Att}
\bibfield{author}{\bibinfo{person}{Jan~K Chorowski}, \bibinfo{person}{Dzmitry
  Bahdanau}, \bibinfo{person}{Dmitriy Serdyuk}, \bibinfo{person}{Kyunghyun
  Cho}, {and} \bibinfo{person}{Yoshua Bengio}.}
  \bibinfo{year}{2015}\natexlab{}.
\newblock \showarticletitle{Attention-Based Models for Speech Recognition}.
\newblock In \bibinfo{booktitle}{\emph{Advances in Neural Information
  Processing Systems 28}}, \bibfield{editor}{\bibinfo{person}{C.~Cortes},
  \bibinfo{person}{N.~D. Lawrence}, \bibinfo{person}{D.~D. Lee},
  \bibinfo{person}{M.~Sugiyama}, {and} \bibinfo{person}{R.~Garnett}} (Eds.).
  \bibinfo{publisher}{Curran Associates, Inc.}, \bibinfo{pages}{577--585}.
\newblock


\bibitem[\protect\citeauthoryear{Chung, Gulcehre, Cho, and Bengio}{Chung
  et~al\mbox{.}}{2014}]%
        {chung2014empirical}
\bibfield{author}{\bibinfo{person}{Junyoung Chung}, \bibinfo{person}{Caglar
  Gulcehre}, \bibinfo{person}{Kyunghyun Cho}, {and} \bibinfo{person}{Yoshua
  Bengio}.} \bibinfo{year}{2014}\natexlab{}.
\newblock \showarticletitle{Empirical evaluation of gated recurrent neural
  networks on sequence modeling}. In \bibinfo{booktitle}{\emph{NIPS 2014
  Workshop on Deep Learning, December 2014}}.
\newblock


\bibitem[\protect\citeauthoryear{Defferrard, Bresson, and
  Vandergheynst}{Defferrard et~al\mbox{.}}{2016}]%
        {Defferrard2016Convolutional}
\bibfield{author}{\bibinfo{person}{Micha\"{e}l Defferrard},
  \bibinfo{person}{Xavier Bresson}, {and} \bibinfo{person}{Pierre
  Vandergheynst}.} \bibinfo{year}{2016}\natexlab{}.
\newblock \showarticletitle{Convolutional Neural Networks on Graphs with Fast
  Localized Spectral Filtering}.
\newblock In \bibinfo{booktitle}{\emph{Advances in Neural Information
  Processing Systems 29}}, \bibfield{editor}{\bibinfo{person}{D.~D. Lee},
  \bibinfo{person}{M.~Sugiyama}, \bibinfo{person}{U.~V. Luxburg},
  \bibinfo{person}{I.~Guyon}, {and} \bibinfo{person}{R.~Garnett}} (Eds.).
  \bibinfo{publisher}{Curran Associates, Inc.}, \bibinfo{pages}{3844--3852}.
\newblock


\bibitem[\protect\citeauthoryear{Gilmer, Schoenholz, Riley, Vinyals, and
  Dahl}{Gilmer et~al\mbox{.}}{2017}]%
        {Gilmer2017Neural}
\bibfield{author}{\bibinfo{person}{Justin Gilmer}, \bibinfo{person}{Samuel~S.
  Schoenholz}, \bibinfo{person}{Patrick~F. Riley}, \bibinfo{person}{Oriol
  Vinyals}, {and} \bibinfo{person}{George~E. Dahl}.}
  \bibinfo{year}{2017}\natexlab{}.
\newblock \showarticletitle{Neural Message Passing for Quantum Chemistry}
  \emph{(\bibinfo{series}{ICML’17})}. \bibinfo{publisher}{JMLR.org},
  \bibinfo{pages}{1263–1272}.
\newblock


\bibitem[\protect\citeauthoryear{Goldberg}{Goldberg}{2017}]%
        {goldberg2017neural}
\bibfield{author}{\bibinfo{person}{Yoav Goldberg}.}
  \bibinfo{year}{2017}\natexlab{}.
\newblock \showarticletitle{Neural network methods for natural language
  processing}.
\newblock \bibinfo{journal}{\emph{Synthesis Lectures on Human Language
  Technologies}} \bibinfo{volume}{10}, \bibinfo{number}{1}
  (\bibinfo{year}{2017}), \bibinfo{pages}{1--309}.
\newblock


\bibitem[\protect\citeauthoryear{{Gori}, {Monfardini}, and {Scarselli}}{{Gori}
  et~al\mbox{.}}{2005}]%
        {gori2005new}
\bibfield{author}{\bibinfo{person}{M. {Gori}}, \bibinfo{person}{G.
  {Monfardini}}, {and} \bibinfo{person}{F. {Scarselli}}.}
  \bibinfo{year}{2005}\natexlab{}.
\newblock \showarticletitle{A new model for learning in graph domains}. In
  \bibinfo{booktitle}{\emph{Proceedings. 2005 IEEE International Joint
  Conference on Neural Networks, 2005.}}, Vol.~\bibinfo{volume}{2}.
  \bibinfo{pages}{729--734 vol. 2}.
\newblock
\showISSN{2161-4407}


\bibitem[\protect\citeauthoryear{Goyal and Ferrara}{Goyal and Ferrara}{2018}]%
        {GOYAL2018Graph}
\bibfield{author}{\bibinfo{person}{Palash Goyal} {and} \bibinfo{person}{Emilio
  Ferrara}.} \bibinfo{year}{2018}\natexlab{}.
\newblock \showarticletitle{Graph embedding techniques, applications, and
  performance: A survey}.
\newblock \bibinfo{journal}{\emph{nowledge-Based Systems}}
  \bibinfo{volume}{151} (\bibinfo{year}{2018}), \bibinfo{pages}{78 -- 94}.
\newblock
\showISSN{0950-7051}


\bibitem[\protect\citeauthoryear{Hamilton, Ying, and Leskovec}{Hamilton
  et~al\mbox{.}}{2017}]%
        {Hamilton2017Inductive}
\bibfield{author}{\bibinfo{person}{Will Hamilton}, \bibinfo{person}{Zhitao
  Ying}, {and} \bibinfo{person}{Jure Leskovec}.}
  \bibinfo{year}{2017}\natexlab{}.
\newblock \showarticletitle{Inductive Representation Learning on Large Graphs}.
\newblock In \bibinfo{booktitle}{\emph{Advances in Neural Information
  Processing Systems 30}}, \bibfield{editor}{\bibinfo{person}{I.~Guyon},
  \bibinfo{person}{U.~V. Luxburg}, \bibinfo{person}{S.~Bengio},
  \bibinfo{person}{H.~Wallach}, \bibinfo{person}{R.~Fergus},
  \bibinfo{person}{S.~Vishwanathan}, {and} \bibinfo{person}{R.~Garnett}}
  (Eds.). \bibinfo{publisher}{Curran Associates, Inc.},
  \bibinfo{pages}{1024--1034}.
\newblock


\bibitem[\protect\citeauthoryear{{Hershey}, {Chaudhuri}, {Ellis}, {Gemmeke},
  {Jansen}, {Moore}, {Plakal}, {Platt}, {Saurous}, {Seybold}, {Slaney},
  {Weiss}, and {Wilson}}{{Hershey} et~al\mbox{.}}{2017}]%
        {Hershey2017CNN}
\bibfield{author}{\bibinfo{person}{S. {Hershey}}, \bibinfo{person}{S.
  {Chaudhuri}}, \bibinfo{person}{D.~P.~W. {Ellis}}, \bibinfo{person}{J.~F.
  {Gemmeke}}, \bibinfo{person}{A. {Jansen}}, \bibinfo{person}{R.~C. {Moore}},
  \bibinfo{person}{M. {Plakal}}, \bibinfo{person}{D. {Platt}},
  \bibinfo{person}{R.~A. {Saurous}}, \bibinfo{person}{B. {Seybold}},
  \bibinfo{person}{M. {Slaney}}, \bibinfo{person}{R.~J. {Weiss}}, {and}
  \bibinfo{person}{K. {Wilson}}.} \bibinfo{year}{2017}\natexlab{}.
\newblock \showarticletitle{CNN architectures for large-scale audio
  classification}. In \bibinfo{booktitle}{\emph{2017 IEEE International
  Conference on Acoustics, Speech and Signal Processing (ICASSP)}}.
  \bibinfo{pages}{131--135}.
\newblock
\showISSN{2379-190X}


\bibitem[\protect\citeauthoryear{Kipf and Welling}{Kipf and Welling}{2017}]%
        {Kipf2016Semi-Supervised}
\bibfield{author}{\bibinfo{person}{Thomas~N. Kipf} {and} \bibinfo{person}{Max
  Welling}.} \bibinfo{year}{2017}\natexlab{}.
\newblock \showarticletitle{Semi-Supervised Classification with Graph
  Convolutional Networks}.
\newblock  (\bibinfo{date}{April} \bibinfo{year}{2017}).
\newblock


\bibitem[\protect\citeauthoryear{Krizhevsky, Sutskever, and Hinton}{Krizhevsky
  et~al\mbox{.}}{2012}]%
        {Krizhevsky2012ImageNet}
\bibfield{author}{\bibinfo{person}{Alex Krizhevsky}, \bibinfo{person}{Ilya
  Sutskever}, {and} \bibinfo{person}{Geoffrey~E Hinton}.}
  \bibinfo{year}{2012}\natexlab{}.
\newblock \showarticletitle{ImageNet Classification with Deep Convolutional
  Neural Networks}. In \bibinfo{booktitle}{\emph{Advances in Neural Information
  Processing Systems 25}}, \bibfield{editor}{\bibinfo{person}{F.~Pereira},
  \bibinfo{person}{C.~J.~C. Burges}, \bibinfo{person}{L.~Bottou}, {and}
  \bibinfo{person}{K.~Q. Weinberger}} (Eds.). \bibinfo{publisher}{Curran
  Associates, Inc.}, \bibinfo{pages}{1097--1105}.
\newblock


\bibitem[\protect\citeauthoryear{Kumar, Irsoy, Ondruska, Iyyer, Bradbury,
  Gulrajani, Zhong, Paulus, and Socher}{Kumar et~al\mbox{.}}{2016}]%
        {kumar2016ask}
\bibfield{author}{\bibinfo{person}{Ankit Kumar}, \bibinfo{person}{Ozan Irsoy},
  \bibinfo{person}{Peter Ondruska}, \bibinfo{person}{Mohit Iyyer},
  \bibinfo{person}{James Bradbury}, \bibinfo{person}{Ishaan Gulrajani},
  \bibinfo{person}{Victor Zhong}, \bibinfo{person}{Romain Paulus}, {and}
  \bibinfo{person}{Richard Socher}.} \bibinfo{year}{2016}\natexlab{}.
\newblock \showarticletitle{Ask me anything: Dynamic memory networks for
  natural language processing}. In \bibinfo{booktitle}{\emph{International
  conference on machine learning}}. \bibinfo{pages}{1378--1387}.
\newblock


\bibitem[\protect\citeauthoryear{Kurt, Ture, and Kurum}{Kurt
  et~al\mbox{.}}{2008}]%
        {Kurt2008Comparing}
\bibfield{author}{\bibinfo{person}{Imran Kurt}, \bibinfo{person}{Mevlut Ture},
  {and} \bibinfo{person}{A.~Turhan Kurum}.} \bibinfo{year}{2008}\natexlab{}.
\newblock \showarticletitle{Comparing performances of logistic regression,
  classification and regression tree, and neural networks for predicting
  coronary artery disease}.
\newblock \bibinfo{journal}{\emph{Expert Systems with Applications}}
  \bibinfo{volume}{34}, \bibinfo{number}{1} (\bibinfo{year}{2008}),
  \bibinfo{pages}{366 -- 374}.
\newblock
\showISSN{0957-4174}


\bibitem[\protect\citeauthoryear{Li, Tarlow, Brockschmidt, and Zemel}{Li
  et~al\mbox{.}}{2016}]%
        {li2015gated}
\bibfield{author}{\bibinfo{person}{Yujia Li}, \bibinfo{person}{Daniel Tarlow},
  \bibinfo{person}{Marc Brockschmidt}, {and} \bibinfo{person}{Richard~S.
  Zemel}.} \bibinfo{year}{2016}\natexlab{}.
\newblock \showarticletitle{Gated Graph Sequence Neural Networks}. In
  \bibinfo{booktitle}{\emph{4th International Conference on Learning
  Representations, {ICLR} 2016, San Juan, Puerto Rico, May 2-4, 2016,
  Conference Track Proceedings}}.
\newblock


\bibitem[\protect\citeauthoryear{Litwin and Stoeckel}{Litwin and
  Stoeckel}{2016}]%
        {Litwin2016Social}
\bibfield{author}{\bibinfo{person}{Howard Litwin} {and}
  \bibinfo{person}{Kimberly~J. Stoeckel}.} \bibinfo{year}{2016}\natexlab{}.
\newblock \showarticletitle{Social Network, Activity Participation, and
  Cognition: A Complex Relationship}.
\newblock \bibinfo{journal}{\emph{Research on Aging}} \bibinfo{volume}{38},
  \bibinfo{number}{1} (\bibinfo{year}{2016}), \bibinfo{pages}{76--97}.
\newblock
\newblock
\shownote{PMID: 25878191.}


\bibitem[\protect\citeauthoryear{{Micheli}}{{Micheli}}{2009}]%
        {Micheli2009Neural}
\bibfield{author}{\bibinfo{person}{A. {Micheli}}.}
  \bibinfo{year}{2009}\natexlab{}.
\newblock \showarticletitle{Neural Network for Graphs: A Contextual
  Constructive Approach}.
\newblock \bibinfo{journal}{\emph{IEEE Transactions on Neural Networks}}
  \bibinfo{volume}{20}, \bibinfo{number}{3} (\bibinfo{date}{March}
  \bibinfo{year}{2009}), \bibinfo{pages}{498--511}.
\newblock
\showISSN{1941-0093}


\bibitem[\protect\citeauthoryear{Milo, Shen-Orr, Itzkovitz, Kashtan,
  Chklovskii, and Alon}{Milo et~al\mbox{.}}{2002}]%
        {Milo2002Motif}
\bibfield{author}{\bibinfo{person}{R. Milo}, \bibinfo{person}{S. Shen-Orr},
  \bibinfo{person}{S. Itzkovitz}, \bibinfo{person}{N. Kashtan},
  \bibinfo{person}{D. Chklovskii}, {and} \bibinfo{person}{U. Alon}.}
  \bibinfo{year}{2002}\natexlab{}.
\newblock \showarticletitle{Network Motifs: Simple Building Blocks of Complex
  Networks}.
\newblock \bibinfo{journal}{\emph{Science}} \bibinfo{volume}{298},
  \bibinfo{number}{5594} (\bibinfo{year}{2002}), \bibinfo{pages}{824--827}.
\newblock
\showISSN{0036-8075}


\bibitem[\protect\citeauthoryear{Nguyen, Do, Calderbank, and
  Deligiannis}{Nguyen et~al\mbox{.}}{2019}]%
        {nguyen2019fake}
\bibfield{author}{\bibinfo{person}{Duc~Minh Nguyen}, \bibinfo{person}{Tien~Huu
  Do}, \bibinfo{person}{Robert Calderbank}, {and} \bibinfo{person}{Nikos
  Deligiannis}.} \bibinfo{year}{2019}\natexlab{}.
\newblock \showarticletitle{Fake news detection using deep markov random
  fields}. In \bibinfo{booktitle}{\emph{Proceedings of the 2019 Conference of
  the North American Chapter of the Association for Computational Linguistics:
  Human Language Technologies, Volume 1 (Long and Short Papers)}}.
  \bibinfo{pages}{1391--1400}.
\newblock


\bibitem[\protect\citeauthoryear{{Ortega}, {Frossard}, {Kovačević}, {Moura},
  and {Vandergheynst}}{{Ortega} et~al\mbox{.}}{2018}]%
        {Ortega2018GSP}
\bibfield{author}{\bibinfo{person}{A. {Ortega}}, \bibinfo{person}{P.
  {Frossard}}, \bibinfo{person}{J. {Kovačević}}, \bibinfo{person}{J.~M.~F.
  {Moura}}, {and} \bibinfo{person}{P. {Vandergheynst}}.}
  \bibinfo{year}{2018}\natexlab{}.
\newblock \showarticletitle{Graph Signal Processing: Overview, Challenges, and
  Applications}.
\newblock \bibinfo{journal}{\emph{Proc. IEEE}} \bibinfo{volume}{106},
  \bibinfo{number}{5} (\bibinfo{date}{May} \bibinfo{year}{2018}),
  \bibinfo{pages}{808--828}.
\newblock
\showISSN{1558-2256}


\bibitem[\protect\citeauthoryear{{Perraudin} and {Vandergheynst}}{{Perraudin}
  and {Vandergheynst}}{2017}]%
        {Perraudin2017Stationary}
\bibfield{author}{\bibinfo{person}{N. {Perraudin}} {and} \bibinfo{person}{P.
  {Vandergheynst}}.} \bibinfo{year}{2017}\natexlab{}.
\newblock \showarticletitle{Stationary Signal Processing on Graphs}.
\newblock \bibinfo{journal}{\emph{IEEE Transactions on Signal Processing}}
  \bibinfo{volume}{65}, \bibinfo{number}{13} (\bibinfo{date}{July}
  \bibinfo{year}{2017}), \bibinfo{pages}{3462--3477}.
\newblock
\showISSN{1941-0476}


\bibitem[\protect\citeauthoryear{Qiao, Zhang, Chen, and Shen}{Qiao
  et~al\mbox{.}}{2018}]%
        {QIAO2018Data}
\bibfield{author}{\bibinfo{person}{Lishan Qiao}, \bibinfo{person}{Limei Zhang},
  \bibinfo{person}{Songcan Chen}, {and} \bibinfo{person}{Dinggang Shen}.}
  \bibinfo{year}{2018}\natexlab{}.
\newblock \showarticletitle{Data-driven graph construction and graph learning:
  A review}.
\newblock \bibinfo{journal}{\emph{Neurocomputing}}  \bibinfo{volume}{312}
  (\bibinfo{year}{2018}), \bibinfo{pages}{336 -- 351}.
\newblock
\showISSN{0925-2312}


\bibitem[\protect\citeauthoryear{Rossi, Ahmed, and Koh}{Rossi
  et~al\mbox{.}}{2018}]%
        {rossi2018higher}
\bibfield{author}{\bibinfo{person}{Ryan~A Rossi}, \bibinfo{person}{Nesreen~K
  Ahmed}, {and} \bibinfo{person}{Eunyee Koh}.} \bibinfo{year}{2018}\natexlab{}.
\newblock \showarticletitle{Higher-order network representation learning}. In
  \bibinfo{booktitle}{\emph{Companion Proceedings of the The Web Conference
  2018}}. International World Wide Web Conferences Steering Committee,
  \bibinfo{pages}{3--4}.
\newblock


\bibitem[\protect\citeauthoryear{{Rossi}, {Zhou}, and {Ahmed}}{{Rossi}
  et~al\mbox{.}}{2018}]%
        {Rossi2018Deep}
\bibfield{author}{\bibinfo{person}{R.~A. {Rossi}}, \bibinfo{person}{R. {Zhou}},
  {and} \bibinfo{person}{N. {Ahmed}}.} \bibinfo{year}{2018}\natexlab{}.
\newblock \showarticletitle{Deep Inductive Graph Representation Learning}.
\newblock \bibinfo{journal}{\emph{IEEE Transactions on Knowledge and Data
  Engineering}} (\bibinfo{year}{2018}), \bibinfo{pages}{1--1}.
\newblock
\showISSN{2326-3865}


\bibitem[\protect\citeauthoryear{Sak, Senior, and Beaufays}{Sak
  et~al\mbox{.}}{2014}]%
        {sak2014long}
\bibfield{author}{\bibinfo{person}{Ha{\c{s}}im Sak}, \bibinfo{person}{Andrew
  Senior}, {and} \bibinfo{person}{Fran{\c{c}}oise Beaufays}.}
  \bibinfo{year}{2014}\natexlab{}.
\newblock \showarticletitle{Long short-term memory recurrent neural network
  architectures for large scale acoustic modeling}. In
  \bibinfo{booktitle}{\emph{Fifteenth annual conference of the international
  speech communication association}}.
\newblock


\bibitem[\protect\citeauthoryear{{Sandryhaila} and {Moura}}{{Sandryhaila} and
  {Moura}}{2013}]%
        {Sandryhaila2013Discrete}
\bibfield{author}{\bibinfo{person}{A. {Sandryhaila}} {and}
  \bibinfo{person}{J.~M.~F. {Moura}}.} \bibinfo{year}{2013}\natexlab{}.
\newblock \showarticletitle{Discrete signal processing on graphs: Graph fourier
  transform}. In \bibinfo{booktitle}{\emph{2013 IEEE International Conference
  on Acoustics, Speech and Signal Processing}}. \bibinfo{pages}{6167--6170}.
\newblock
\showISSN{2379-190X}


\bibitem[\protect\citeauthoryear{{Scarselli}, {Gori}, {Tsoi}, {Hagenbuchner},
  and {Monfardini}}{{Scarselli} et~al\mbox{.}}{2009}]%
        {scarselli2008graph}
\bibfield{author}{\bibinfo{person}{F. {Scarselli}}, \bibinfo{person}{M.
  {Gori}}, \bibinfo{person}{A.~C. {Tsoi}}, \bibinfo{person}{M. {Hagenbuchner}},
  {and} \bibinfo{person}{G. {Monfardini}}.} \bibinfo{year}{2009}\natexlab{}.
\newblock \showarticletitle{The Graph Neural Network Model}.
\newblock \bibinfo{journal}{\emph{IEEE Transactions on Neural Networks}}
  \bibinfo{volume}{20}, \bibinfo{number}{1} (\bibinfo{date}{Jan}
  \bibinfo{year}{2009}), \bibinfo{pages}{61--80}.
\newblock
\showISSN{1941-0093}


\bibitem[\protect\citeauthoryear{Seo, Defferrard, Vandergheynst, and
  Bresson}{Seo et~al\mbox{.}}{2018}]%
        {Seo2018Structured}
\bibfield{author}{\bibinfo{person}{Youngjoo Seo}, \bibinfo{person}{Micha{\"e}l
  Defferrard}, \bibinfo{person}{Pierre Vandergheynst}, {and}
  \bibinfo{person}{Xavier Bresson}.} \bibinfo{year}{2018}\natexlab{}.
\newblock \showarticletitle{Structured Sequence Modeling with Graph
  Convolutional Recurrent Networks}. In \bibinfo{booktitle}{\emph{Neural
  Information Processing}}, \bibfield{editor}{\bibinfo{person}{Long Cheng},
  \bibinfo{person}{Andrew Chi~Sing Leung}, {and} \bibinfo{person}{Seiichi
  Ozawa}} (Eds.). \bibinfo{publisher}{Springer International Publishing},
  \bibinfo{address}{Cham}, \bibinfo{pages}{362--373}.
\newblock
\showISBNx{978-3-030-04167-0}


\bibitem[\protect\citeauthoryear{{Singh}, {Thakur}, and {Sharma}}{{Singh}
  et~al\mbox{.}}{2016}]%
        {Singh2016review}
\bibfield{author}{\bibinfo{person}{A. {Singh}}, \bibinfo{person}{N. {Thakur}},
  {and} \bibinfo{person}{A. {Sharma}}.} \bibinfo{year}{2016}\natexlab{}.
\newblock \showarticletitle{A review of supervised machine learning
  algorithms}. In \bibinfo{booktitle}{\emph{2016 3rd International Conference
  on Computing for Sustainable Global Development (INDIACom)}}.
  \bibinfo{pages}{1310--1315}.
\newblock
\showISSN{null}


\bibitem[\protect\citeauthoryear{Velickovic, Cucurull, Casanova, Romero,
  Li{\`{o}}, and Bengio}{Velickovic et~al\mbox{.}}{2018}]%
        {Velickovic2018Graph}
\bibfield{author}{\bibinfo{person}{Petar Velickovic}, \bibinfo{person}{Guillem
  Cucurull}, \bibinfo{person}{Arantxa Casanova}, \bibinfo{person}{Adriana
  Romero}, \bibinfo{person}{Pietro Li{\`{o}}}, {and} \bibinfo{person}{Yoshua
  Bengio}.} \bibinfo{year}{2018}\natexlab{}.
\newblock \showarticletitle{Graph Attention Networks}. In
  \bibinfo{booktitle}{\emph{6th International Conference on Learning
  Representations, {ICLR} 2018, Vancouver, BC, Canada, April 30 - May 3, 2018,
  Conference Track Proceedings}}.
\newblock


\bibitem[\protect\citeauthoryear{Vorgia, Triantafyllou, and Koulouris}{Vorgia
  et~al\mbox{.}}{2017}]%
        {Vorgia2017Hypatia}
\bibfield{author}{\bibinfo{person}{Frosso Vorgia}, \bibinfo{person}{Ioannis
  Triantafyllou}, {and} \bibinfo{person}{Alexandros Koulouris}.}
  \bibinfo{year}{2017}\natexlab{}.
\newblock \showarticletitle{Hypatia Digital Library: A Text Classification
  Approach Based on Abstracts}. In \bibinfo{booktitle}{\emph{Strategic
  Innovative Marketing}}, \bibfield{editor}{\bibinfo{person}{Androniki
  Kavoura}, \bibinfo{person}{Damianos~P. Sakas}, {and} \bibinfo{person}{Petros
  Tomaras}} (Eds.). \bibinfo{publisher}{Springer International Publishing},
  \bibinfo{address}{Cham}, \bibinfo{pages}{727--733}.
\newblock


\bibitem[\protect\citeauthoryear{{Vu}, {Gupta}, {Adel}, and {Schütze}}{{Vu}
  et~al\mbox{.}}{2016}]%
        {Vu2016Bi}
\bibfield{author}{\bibinfo{person}{N.~T. {Vu}}, \bibinfo{person}{P. {Gupta}},
  \bibinfo{person}{H. {Adel}}, {and} \bibinfo{person}{H. {Schütze}}.}
  \bibinfo{year}{2016}\natexlab{}.
\newblock \showarticletitle{Bi-directional recurrent neural network with
  ranking loss for spoken language understanding}. In
  \bibinfo{booktitle}{\emph{2016 IEEE International Conference on Acoustics,
  Speech and Signal Processing (ICASSP)}}. \bibinfo{pages}{6060--6064}.
\newblock
\showISSN{2379-190X}


\bibitem[\protect\citeauthoryear{{Wang}, {Liu}, {Yang}, {Kong}, and
  {Xia}}{{Wang} et~al\mbox{.}}{2019}]%
        {Wang2019Sustainable}
\bibfield{author}{\bibinfo{person}{W. {Wang}}, \bibinfo{person}{J. {Liu}},
  \bibinfo{person}{Z. {Yang}}, \bibinfo{person}{X. {Kong}}, {and}
  \bibinfo{person}{F. {Xia}}.} \bibinfo{year}{2019}\natexlab{}.
\newblock \showarticletitle{Sustainable Collaborator Recommendation Based on
  Conference Closure}.
\newblock \bibinfo{journal}{\emph{IEEE Transactions on Computational Social
  Systems}} \bibinfo{volume}{6}, \bibinfo{number}{2} (\bibinfo{date}{April}
  \bibinfo{year}{2019}), \bibinfo{pages}{311--322}.
\newblock
\showISSN{2373-7476}


\bibitem[\protect\citeauthoryear{White, Currie, and Lockett}{White
  et~al\mbox{.}}{2016}]%
        {white2016pluralized}
\bibfield{author}{\bibinfo{person}{Leroy White}, \bibinfo{person}{Graeme
  Currie}, {and} \bibinfo{person}{Andy Lockett}.}
  \bibinfo{year}{2016}\natexlab{}.
\newblock \showarticletitle{Pluralized leadership in complex organizations:
  Exploring the cross network effects between formal and informal leadership
  relations}.
\newblock \bibinfo{journal}{\emph{The Leadership Quarterly}}
  \bibinfo{volume}{27}, \bibinfo{number}{2} (\bibinfo{year}{2016}),
  \bibinfo{pages}{280--297}.
\newblock


\bibitem[\protect\citeauthoryear{Wu, Williams, Chen, Khabsa, Caragea, Tuarob,
  Ororbia, Jordan, Mitra, and Giles}{Wu et~al\mbox{.}}{2015}]%
        {wu2015citeseerx}
\bibfield{author}{\bibinfo{person}{Jian Wu}, \bibinfo{person}{Kyle~Mark
  Williams}, \bibinfo{person}{Hung-Hsuan Chen}, \bibinfo{person}{Madian
  Khabsa}, \bibinfo{person}{Cornelia Caragea}, \bibinfo{person}{Suppawong
  Tuarob}, \bibinfo{person}{Alexander~G Ororbia}, \bibinfo{person}{Douglas
  Jordan}, \bibinfo{person}{Prasenjit Mitra}, {and} \bibinfo{person}{C~Lee
  Giles}.} \bibinfo{year}{2015}\natexlab{}.
\newblock \showarticletitle{Citeseerx: Ai in a digital library search engine}.
\newblock \bibinfo{journal}{\emph{AI Magazine}} \bibinfo{volume}{36},
  \bibinfo{number}{3} (\bibinfo{year}{2015}), \bibinfo{pages}{35--48}.
\newblock


\bibitem[\protect\citeauthoryear{Xiong, Xiao, Cao, Gong, Fang, and Zhou}{Xiong
  et~al\mbox{.}}{2019}]%
        {xiong2019good}
\bibfield{author}{\bibinfo{person}{Fu Xiong}, \bibinfo{person}{Yang Xiao},
  \bibinfo{person}{Zhiguo Cao}, \bibinfo{person}{Kaicheng Gong},
  \bibinfo{person}{Zhiwen Fang}, {and} \bibinfo{person}{Joey~Tianyi Zhou}.}
  \bibinfo{year}{2019}\natexlab{}.
\newblock \showarticletitle{Good practices on building effective CNN baseline
  model for person re-identification}. In \bibinfo{booktitle}{\emph{Tenth
  International Conference on Graphics and Image Processing (ICGIP 2018)}},
  Vol.~\bibinfo{volume}{11069}. International Society for Optics and Photonics,
  \bibinfo{pages}{110690I}.
\newblock


\bibitem[\protect\citeauthoryear{Xu, Dong, Luo, and Wang}{Xu
  et~al\mbox{.}}{2018}]%
        {xu2018research}
\bibfield{author}{\bibinfo{person}{Haiyun Xu}, \bibinfo{person}{Kun Dong},
  \bibinfo{person}{Rui Luo}, {and} \bibinfo{person}{Chao Wang}.}
  \bibinfo{year}{2018}\natexlab{}.
\newblock \showarticletitle{Research on Topic Recognition Based on Multivariate
  Relation Fusion}. In \bibinfo{booktitle}{\emph{23rd International Conference
  on Science and Technology Indicators (STI 2018), September 12-14, 2018,
  Leiden, The Netherlands}}. Centre for Science and Technology Studies (CWTS).
\newblock


\bibitem[\protect\citeauthoryear{{Yu}, {Xia}, and {Liu}}{{Yu}
  et~al\mbox{.}}{2019a}]%
        {Yu2019Academic}
\bibfield{author}{\bibinfo{person}{S. {Yu}}, \bibinfo{person}{F. {Xia}}, {and}
  \bibinfo{person}{H. {Liu}}.} \bibinfo{year}{2019}\natexlab{a}.
\newblock \showarticletitle{Academic Team Formulation Based on Liebig’s
  Barrel: Discovery of Anticask Effect}.
\newblock \bibinfo{journal}{\emph{IEEE Transactions on Computational Social
  Systems}} \bibinfo{volume}{6}, \bibinfo{number}{5} (\bibinfo{date}{Oct}
  \bibinfo{year}{2019}), \bibinfo{pages}{1083--1094}.
\newblock
\showISSN{2373-7476}


\bibitem[\protect\citeauthoryear{{Yu}, {Xia}, {Zhang}, {Ning}, {Zhong}, and
  {Liu}}{{Yu} et~al\mbox{.}}{2017}]%
        {Yu2017Team}
\bibfield{author}{\bibinfo{person}{S. {Yu}}, \bibinfo{person}{F. {Xia}},
  \bibinfo{person}{K. {Zhang}}, \bibinfo{person}{Z. {Ning}},
  \bibinfo{person}{J. {Zhong}}, {and} \bibinfo{person}{C. {Liu}}.}
  \bibinfo{year}{2017}\natexlab{}.
\newblock \showarticletitle{Team Recognition in Big Scholarly Data: Exploring
  Collaboration Intensity}. In \bibinfo{booktitle}{\emph{3rd Intl Conf on Big
  Data Intelligence and Computing(DataCom)}}. \bibinfo{pages}{925--932}.
\newblock
\showISSN{null}


\bibitem[\protect\citeauthoryear{{Yu}, {Xu}, {Zhang}, {Xia}, {Almakhadmeh}, and
  {Tolba}}{{Yu} et~al\mbox{.}}{2019b}]%
        {Yu2019Motifs}
\bibfield{author}{\bibinfo{person}{S. {Yu}}, \bibinfo{person}{J. {Xu}},
  \bibinfo{person}{C. {Zhang}}, \bibinfo{person}{F. {Xia}}, \bibinfo{person}{Z.
  {Almakhadmeh}}, {and} \bibinfo{person}{A. {Tolba}}.}
  \bibinfo{year}{2019}\natexlab{b}.
\newblock \showarticletitle{Motifs in Big Networks: Methods and Applications}.
\newblock \bibinfo{journal}{\emph{IEEE Access}}  \bibinfo{volume}{7}
  (\bibinfo{year}{2019}), \bibinfo{pages}{183322--183338}.
\newblock
\showISSN{2169-3536}


\bibitem[\protect\citeauthoryear{Zhang, Zuo, Gu, and Zhang}{Zhang
  et~al\mbox{.}}{2017}]%
        {Zhang2017Learning}
\bibfield{author}{\bibinfo{person}{Kai Zhang}, \bibinfo{person}{Wangmeng Zuo},
  \bibinfo{person}{Shuhang Gu}, {and} \bibinfo{person}{Lei Zhang}.}
  \bibinfo{year}{2017}\natexlab{}.
\newblock \showarticletitle{Learning Deep CNN Denoiser Prior for Image
  Restoration}. In \bibinfo{booktitle}{\emph{The IEEE Conference on Computer
  Vision and Pattern Recognition (CVPR)}}.
\newblock


\bibitem[\protect\citeauthoryear{Zhang, Zhang, Zeng, Dong, Chen, Huang, Zhu,
  Xu, Cheng, Hou, Cao, and Fan}{Zhang et~al\mbox{.}}{2018}]%
        {zhang2018Multivariate}
\bibfield{author}{\bibinfo{person}{Lihua Zhang}, \bibinfo{person}{Jiachao
  Zhang}, \bibinfo{person}{Guangming Zeng}, \bibinfo{person}{Haoran Dong},
  \bibinfo{person}{Yaoning Chen}, \bibinfo{person}{Chao Huang},
  \bibinfo{person}{Yuan Zhu}, \bibinfo{person}{Rui Xu}, \bibinfo{person}{Yujun
  Cheng}, \bibinfo{person}{Kunjie Hou}, \bibinfo{person}{Weicheng Cao}, {and}
  \bibinfo{person}{Wei Fan}.} \bibinfo{year}{2018}\natexlab{}.
\newblock \showarticletitle{Multivariate relationships between microbial
  communities and environmental variables during co-composting of sewage sludge
  and agricultural waste in the presence of PVP-AgNPs}.
\newblock \bibinfo{journal}{\emph{Bioresource Technology}}
  \bibinfo{volume}{261} (\bibinfo{year}{2018}), \bibinfo{pages}{10 -- 18}.
\newblock
\showISSN{0960-8524}


\end{thebibliography}

\end{document}